\PassOptionsToPackage{square,numbers,sort&compress}{natbib}
\documentclass[5p]{elsarticle}
\usepackage[T1]{fontenc}
\usepackage[utf8]{inputenc}
\usepackage{natbib}
\usepackage{amsmath,amssymb,amsfonts}
\usepackage{graphicx}
\usepackage{xcolor}

\begin{document}

\title{The localization regime in a nutshell}

\author{Nicolas Moutal}
\address{Laboratoire de Physique de la Mati\`{e}re Condens\'{e}e (UMR 7643), \\ 
CNRS -- Ecole Polytechnique, IP Paris, 91128 Palaiseau, France}

\author{Denis~S.~Grebenkov}
	\ead{denis.grebenkov@polytechnique.edu}
\address{Laboratoire de Physique de la Mati\`{e}re Condens\'{e}e (UMR 7643), \\ 
CNRS -- Ecole Polytechnique, IP Paris, 91128 Palaiseau, France}
\address{Institute for Physics and Astronomy, University of Potsdam, 14476 Potsdam-Golm,
Germany}

\begin{abstract}
High diffusion-sensitizing magnetic field gradients have been more and
more often applied nowadays to achieve a better characterization of
the microstructure.  As the resulting spin-echo signal significantly
deviates from the conventional Gaussian form, various models have been
employed to interpret these deviations and to relate them with the
microstructural properties of a sample.  In this paper, we argue that
the non-Gaussian behavior of the signal is a generic universal feature
of the Bloch-Torrey equation.  We provide a simple yet rigorous
description of the localization regime emerging at high extended
gradients and identify its origin as a symmetry breaking at the
reflecting boundary.  We compare the consequent non-Gaussian signal
decay to other diffusion NMR regimes such as slow-diffusion,
motional-narrowing and diffusion-diffraction regimes.  We emphasize
limitations of conventional perturbative techniques and advocate for
non-perturbative approaches which may pave a way to new imaging
modalities in this field.
\end{abstract}


\date{\today}

\maketitle

\section{Introduction}

\begin{figure}[t!]
\centering
\includegraphics[width=0.9\linewidth]{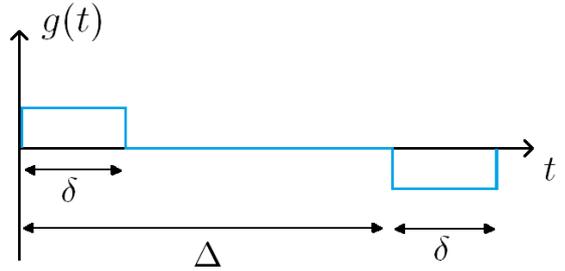} 
\caption{
Schematic representation of the pulsed-gradient spin-echo (PGSE)
sequence with two rectangular gradient pulses of amplitude $g$ and
conventional notations $\delta$ and $\Delta$ for the pulse and
inter-pulse durations \cite{Stejskal1965b,Tanner1968a}.  The second
negative gradient pulse accounts for the refocusing effect of a
$180^\circ$ rf pulse.  While the paper is mainly focused on the
particular case $\Delta = \delta$, the localization regime can also be
observed and analyzed for $\Delta > \delta$ given that the pulse
duration $\delta$ is long enough \cite{Moutal2019b}.  In
Sec. \ref{sec:narrow}, the localization regime is also compared with
the opposite setting of narrow gradient pulses ($\delta \to 0$).}
\label{fig:PGSE}
\end{figure}

After the very first spin echo produced by E. Hahn in 1950
\cite{Hahn1950a}, the NMR has achieved remarkable advances and found
countless applications in physics, chemistry, material sciences,
neurosciences and medicine
\cite{Callaghan1991a,Price2009a,Jones2011a,Tuch2003a,Frahm2004a,LeBihan2012a,Grebenkov2007a,Kiselev2017a,Novikov2018a}.
Such long and intensive developments over seven decades, as well as
spreading into various disciplines, led to some dogmatic views whose
origins are often forgotten or even unknown.  In diffusion NMR, such a
dogma is a {\it perturbative} approach to the study of the
Bloch-Torrey equation and to the consequent analysis of the
macroscopic spin-echo signal.  The Bloch-Torrey equation governs the
time evolution of the transverse magnetization $m(t,\mathbf{r})$ of
the nuclei, from the exciting $90^\circ$ rf pulse till the spin-echo
formation \cite{Torrey1956a}:
\begin{equation}  \label{eq:BT}
\partial_t m(t,\mathbf{r}) = D_0 \nabla^2 m(t,\mathbf{r}) - i \gamma (\mathbf{g}(t) \cdot \mathbf{r}) m(t,\mathbf{r})\, ,
\end{equation}
where $\gamma$ and $D_0$ are the gyromagnetic ratio and the diffusion
coefficient of the nuclei, and $\mathbf{g}(t)$ is the gradient profile
that accounts for the effect of the refocusing $180^\circ$ rf pulse
(Fig. \ref{fig:PGSE}).  Despite the linear form of this partial
differential equation, its exact solution gets a simple closed form
only for free diffusion, from which the signal attenuation follows as
\begin{equation}  \label{eq:S_free0}
S = \exp(-D_0 b),
\end{equation}
where $b \propto g^2$ incorporates the gradient sequence
$\mathbf{g}(t)$ in a standard explicit way \cite{Stejskal1965a}.  This
remarkably simple relation stands at the origin of diffusion NMR:
changing the gradient sequence ($b$) and measuring the resulting
signal ($S$), one accesses the dynamics of the nuclei ($D_0$)
\cite{Douglass1958a}.  Unfortunately, the free diffusion is the only
known setting for which an exact and simple expression for the signal
is available.  In presence of {\it any} microstructure, even for
one-dimensional domains such as a half-line or an interval, the exact
solution of the Bloch-Torrey equation and the consequent signal get a
sophisticated form \cite{Grebenkov2007a,Stoller1991a}.  It is thus not
surprising that most theoretical efforts in the past were dedicated to
obtaining various perturbative approximations for the signal that
could allow to fit and to interpret the measured signal in biological
or mineral samples.  The simplest and the most broadly used one is the
Gaussian phase approximation, in which the microstructure is supposed
to effectively slow down diffusion and thus to reduce the diffusion
coefficient $D_0$.  The signal keeps thus the monoexponential form,
\begin{equation}  \label{eq:GPA}
S \simeq \exp(-D b) , 
\end{equation}
where $D$ is the effective or apparent diffusion coefficient (ADC)
\cite{Woessner1963a,Wayne1966a}.  As the reduction of $D_0$ to $D$ is
caused by the microstructure, an estimation of $D$ from the measured
signal allows one to probe some microstructural properties.  The
estimated ADC can either be directly used as a biomarker of some
pathology in the tissue (e.g., the emphysema in the lungs or a tumor
in the brain
\cite{Tuch2003a,Frahm2004a,LeBihan2012a,vanBeek2004a}) or as an
intermediate quantity for further theoretical interpretations, e.g.,
in the short-time \cite{Mitra1992a,Mitra1993a,Moutal2019a} or
long-time regimes
\cite{Robertson1966a,Wayne1966a,Neuman1974a,Novikov2011a,Novikov2014a}.
The major part of the literature, both experimental and theoretical,
focuses on ADC or its variants \cite{Basser2002a}, sometimes with
abuse \cite{Grebenkov2010a}.

At the same time, practically any diffusion NMR measurement realized
today would reveal deviations from the monoexponential form of the
signal at moderately large $b$-values.  Different improvements have
been proposed to capture these deviations: (i) the bi-exponential
model with two ADCs aiming to characterize two isolated
``compartments'' (e.g., intracellular and extracellular water)
\cite{Niendorf1996a,Mulkern1999a,Clark2000a,Kiselev2007a}; (ii) the
K\"arger model accounting for the exchange between two compartments
\cite{Karger1985a,Karger1988a,Fieremans2010a,Moutal2018a}; (iii) the
distributed model, in which a variety of compartments is represented
by the distribution of ADCs \cite{Pfeuffer1999a,Yablonskiy2003a}; (iv)
the cylinder model \cite{Callaghan1979a,Yablonskiy2002a}, which
accounts for fiber-like anisotropy and averages over random
orientations; (v) anomalous diffusion models, in which the
microstructure severely affects diffusion
\cite{Magin2008a,Palombo2011a}; and (vi) the kurtosis correction,
which stems from the cumulant expansion
\cite{Jensen2005a,Trampel2006a,Kiselev2010a}.  Without discussing
their advantages and drawbacks (see an overview in
\cite{Grebenkov2016a}), we emphasize that all the improvements from
(i) to (v) just ``decorate'' the monoexponential form (\ref{eq:GPA}),
rendering the signal dependence on $b$ more sophisticated but keeping
the essence of the perturbative approach.  The kurtosis correction
makes the first step beyond the Gaussian phase approximation but still
remains perturbative in its nature.  The very possibility of treating
the gradient encoding term in Eq. (\ref{eq:BT}) as a perturbation to
the diffusion operator $D_0\nabla^2$, is one of the key dogmas in the
current theory of diffusion NMR.

In 1991, Stoller, Happer and Dyson have solved exactly the
Bloch-Torrey equation (\ref{eq:BT}) in one dimension and predicted the
emergence of the localization regime, in which the signal decays much
slower at high extended gradient pulses \cite{Stoller1991a}
\begin{equation}  \label{eq:Sloc}
-\log S \propto g^{2/3}  \;.
\end{equation}
This first non-perturbative approach to the Bloch-Torrey equation was
later extended by de Swiet and Sen \cite{deSwiet1994a} and validated
experimentally by H\"urlimann {\it et al.} \cite{Hurlimann1995a}.  The
mathematical complexity of the seminal paper \cite{Stoller1991a} and
the unusual, non-intuitive behavior of the signal led to a common view
onto the localization regime as a sort of pathologic anomalous
exception.  Over many years, these three papers remained under-cited
and largely ignored.  Only recently, the interest to the localization
regime has been revived.  The recent works have shown that, as opposed
to a common belief, the localization regime is not an exception, but a
{\it universal} mathematical feature of the Bloch-Torrey equation
\cite{Grebenkov2014b,Herberthson2017a,Grebenkov2017a,Grebenkov2018b,Almog2018a,Almog2019a,Moutal2019b,Moutal2020a}.
Moreover, the high sensitivity of the signal to the microstructure at
strong gradients presents an unexplored opportunity for new imaging
modalities \cite{Grebenkov2018a}.  Yet, the lack of simple, intuitive
description of the localization regime may present a severe obstacle
for these exciting developments.

In this paper, we fill this gap and provide a relatively simple yet
rigorous explanation of the localization regime and its fascinating
properties.  We also discuss limitations of earlier proposed
hand-waving arguments employed to explain the localization regime.
After this didactic presentation, we summarize the panorama of
different regimes and their relevance to experiments.  Finally, we
argue on the universal character of the localization regime, urge for
the development of a non-perturbative theory of diffusion NMR, and
speculate about future perspectives.

\section{Relevant length scales of diffusion NMR}

For the sake of simplicity, we will consider the basic Stejskal-Tanner
pulsed-gradient spin-echo sequence \cite{Stejskal1965b,Tanner1968a}
with two rectangular gradient pulses, each of amplitude $g$ and
duration $\delta$, and without inter-pulse time (i.e., $\Delta =
\delta$, see Fig. \ref{fig:PGSE}).  To focus on the effects of
diffusion-sensitizing gradients, we ignore $T_1/T_2$ relaxations,
surface relaxation, permeability, susceptibility-induced internal
gradients, Eddy currents, and other experimental features which
usually superimpose with the considered attenuation mechanism and
further complicate the analysis.  We will comment on them at the end
of the paper.

Following \cite{Hurlimann1995a}, we introduce two length scales in
order to distinguish different diffusion NMR regimes: a diffusion
length $\ell_{\mathrm d} = \sqrt{D_0 t}$ (with $t = 2\delta$) and a
gradient length $\ell_g = D_0^{1/3} G^{-1/3}$, where we set a shortcut
notation $G = \gamma g$ for the gradient of the Larmor frequency (see
\ref{section:scales} for a qualitative explanation of the gradient
length scale).  For free diffusion, there is no other length scale,
and the signal attention must be a function of the ratio
$\ell_{\mathrm d}/\ell_g$.  Indeed, one finds \cite{Carr1954a}
\begin{equation}  \label{eq:S_free}
S = \exp\bigl(- \tfrac{1}{12} D_0 G^2 t^3\bigr) = \exp\bigl( - \tfrac{1}{12} (\ell_{\mathrm d}/\ell_g)^6 \bigr)
\end{equation}
that also justifies the introduction of the $b$-value (here, $b =
\tfrac{1}{12} G^2 t^3$, compare with Eq. (\ref{eq:S_free0})) as a
unique parameter representing the diffusion-encoding sequence.

\begin{figure}[t!]
\centering
\includegraphics[width=0.9\linewidth]{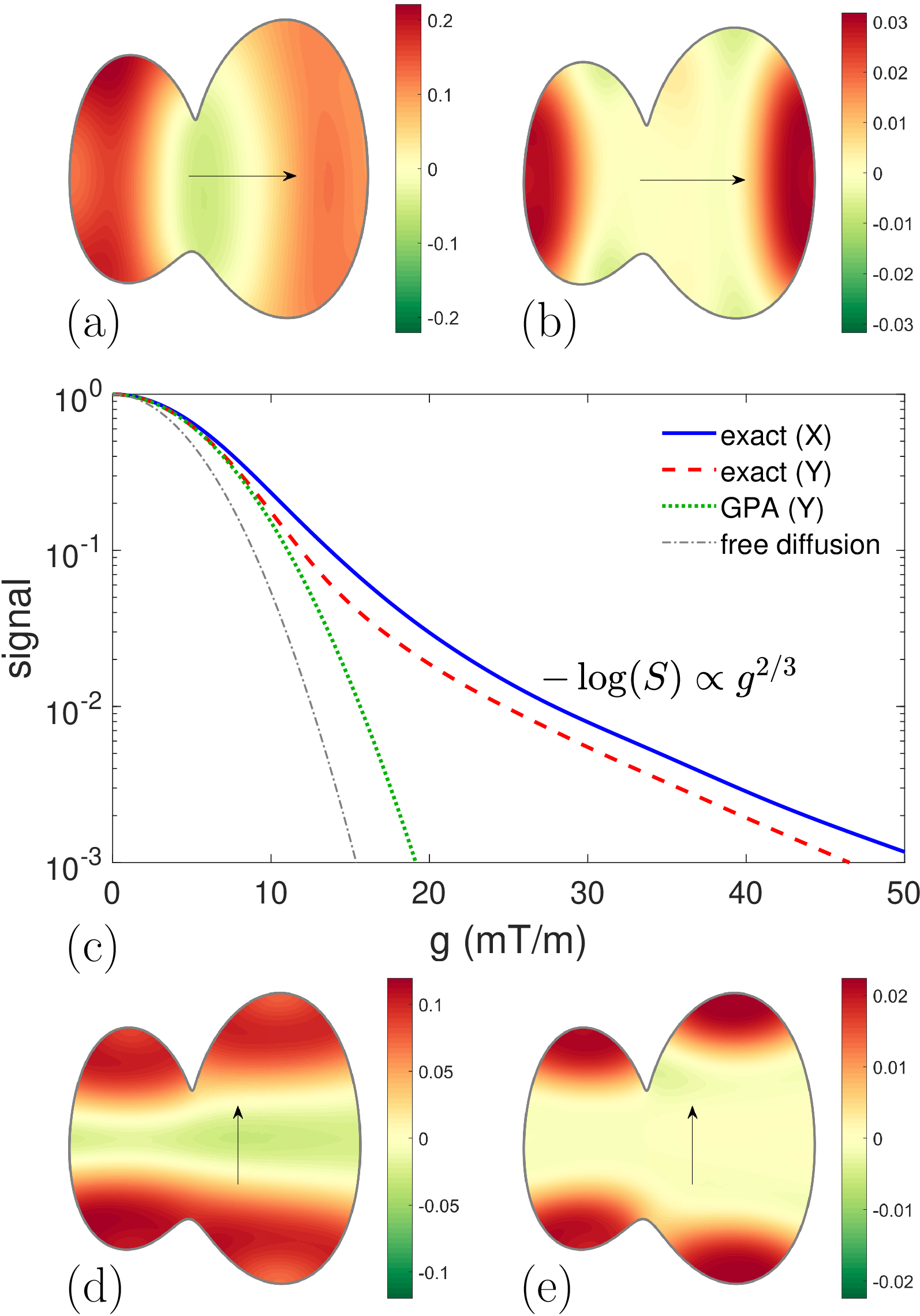}
\caption{
Emergence of the localization regime in a bounded domain with
reflecting boundary filled with xenon gas.  {\bf (a,b,d,e)} The real
part of the transverse magnetization $m(t,\mathbf{r})$ at the echo
time $t = 2\delta$ after the second rectangular gradient pulse shown
on Fig. \ref{fig:PGSE}, with $\Delta = \delta = 6$~ms, $D_0 = 3.7\cdot
10^{-5}$~m$^2$/s, $\gamma = 7.4\cdot 10^7$~s$^{-1}$~T$^{-1}$, and the
linear size of the domain $\ell_{\mathrm s} = 3$~mm.  Panels {\bf
(a,b)} correspond to the horizontally oriented gradient of amplitude
$g = 15$~mT/m {\bf (a)} and $g = 30$~mT/m {\bf (b)}; Panels {\bf
(d,e)} correspond to the vertically oriented gradient of amplitude $g
= 15$~mT/m {\bf (d)} and $g = 30$~mT/m {\bf (e)}.  In this setting,
one has $\ell_{\mathrm d}/\ell_{\mathrm s} = 0.157$, while
$\ell_g/\ell_{\mathrm s} = 0.107$ for $g = 15$~mT/m and
$\ell_g/\ell_{\mathrm s} = 0.085$ for $g = 30$~mT/m.  One clearly
observes the ``pockets'' of localized magnetization. {\bf (c)} The
spin-echo signal $S$ as a function of the gradient.  Solid and dashed
lines show the signals $S_x$ and $S_y$ when the gradient is directed
either along the $x$ axis, or along the $y$ axis, respectively.
Dotted line indicates the Gaussian phase approximation \eqref{eq:GPA}
with $D = 2.39\cdot 10^{-5}$~m$^2$/s obtained from a linear fit of
$-\log(S_y)$ versus $b$-value at small $b = \tfrac{2}{3} \gamma^2 g^2
\delta^3$ (a similar curve with $D = 2.19 \cdot 10^{-5}$~m$^2$/s
obtained from $S_x$ is not shown).  Thin dash-dotted line presents the
free diffusion signal in Eq. \eqref{eq:S_free}.  The signal and the
magnetization profiles were calculated via a matrix formalism based on
the Laplacian eigenmodes computed numerically in Matlab PDEtool, see
\cite{Grebenkov2007a,Grebenkov2008a}.}
\label{fig:magn}
\end{figure}

In turn, the microstructure of a medium (such a brain tissue or a
porous sedimentary rock) is usually incorporated via boundary
conditions to Eq. \eqref{eq:BT} and introduces its own length
scale(s), denoted as $\ell_{\mathrm s}$, resulting in much more
sophisticated dependences of the signal on the experimental parameters
\cite{Song2000a}.
When the gradient amplitude $G$ (or $g$) increases, the gradient
length $\ell_g$ decreases and can eventually become the smallest
length scale of the problem.  In this case, the transverse
magnetization becomes very small everywhere in the bulk, except for a
boundary layer of width $\ell_g$ near the points where the gradient
direction is perpendicular to the boundary (Fig. \ref{fig:magn}).  The
residual magnetization localized in these specific regions produces
the spin-echo signal, which exhibits the ``anomalous'' decay
\eqref{eq:Sloc}, as we discuss below.  While this common description
of the localization regime sounds plausible, an appropriate physical
explanation of this phenomenon is still missing, apart from the
thorough mathematical analysis of the Bloch-Torrey equation.

\section{Why does the magnetization localize?}
\label{section:localization_qualitative}

In this section, we aim at explaining why does the magnetization
localize at high gradients and how does the gradient length $\ell_g$
determine its spatial extent.  We will first revise common
misconceptions and then provide a qualitative explanation for this
behavior.  Then we extend this description to emphasize the difference
between the motional-narrowing and localization regimes.  For clarity,
we consider one-dimensional settings, which can be seen as a zoom of
the local behavior of the magnetization in the orthogonal direction to
the boundary in three dimensions.  This qualitative picture is
justified by the fact that the gradient length $\ell_g$ is supposed to
the smallest scale of the problem.  Even though the exact solution of
the Bloch-Torrey equation in terms of infinite series over Airy
functions is known for one-dimensional domains
\cite{Stoller1991a,Grebenkov2014b}, our goal here is to provide a
simple, intuitively appealing description of the localization
phenomenon.

\subsection{Reduced mean-squared displacement?}

\begin{figure}[t!]
\centering
\includegraphics[width=1.05\linewidth]{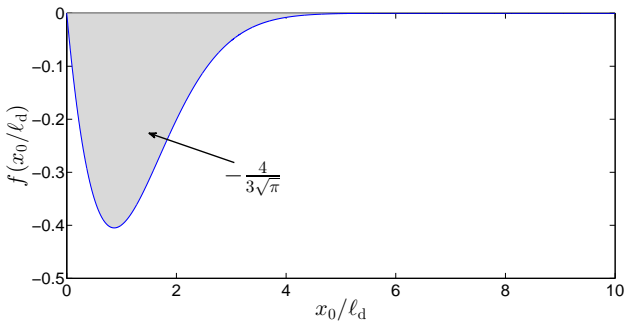} 
\caption{
The correction term $f(x_0/\ell_{\mathrm d})$ given by
Eq. \eqref{eq:u_def} that quantifies the relative decrease in the
one-dimensional mean-squared displacement from
Eq. \eqref{eq:mean_square_displacement_correction_Mitra} due to
reflections on the boundary, as a function of the starting position
$x_0$.  This is also the relative reduction in the effective diffusion
coefficient for particles started from $x_0$.  The shaded area
represents the integral of this correction term over $x_0$ and yields
the numerical prefactor $4/(3\sqrt\pi)$ that was first computed by
Mitra \textit{et al.} in the context of diffusion NMR
\cite{Mitra1992a,Mitra1993a}.  The fact that the integral is finite
expresses that the correction is a boundary effect.}
\label{fig:correction_Mitra}
\end{figure}

The main argument that is commonly put forward to rationalize
localization of the magnetization is that the displacement of
particles along the gradient direction is the most reduced at
boundaries that are perpendicular to the gradient.  Although this
restriction is indeed present, we argue that its effect is far too
weak to explain the drastic change in the signal decay in comparison
to free diffusion.  For a particle diffusing on a half-line
$(0,\infty)$ with reflections at the endpoint $0$, the mean-squared
displacement can be easily found as
\begin{equation}  \label{eq:mean_square_displacement_correction_Mitra}
\mathbb{E}\{(x_t-x_0)^2 \} = 2 \ell_{\mathrm d}^2 \bigl( 1 +  f(x_0/\ell_{\mathrm d}) \bigr),
\end{equation}
where $x_0$ is the starting point, and 
\begin{equation}  \label{eq:u_def}
f(u) = u^2 \left[1 - \mathrm{erf}(u/2)\right] -\frac{2u}{\sqrt\pi}\exp(-u^2/4), 
\end{equation}
with $\mathrm{erf}(z)$ being the Gauss error function.  The correction
term $f(u)$ is illustrated on Fig. \ref{fig:correction_Mitra}, which
reveals that the mean-squared displacement is reduced at most by
$40\%$ of its nominal value $2 \ell_{\mathrm d}^2 = 2D_0 t$ for free
diffusion.  Although this is a strong effect in itself and may result,
for instance, in the edge enhancement \cite{Swiet1995a}, it cannot
explain the ``anomalous'' scaling of the signal in
Eq. (\ref{eq:Sloc}), which is drastically different from the
monoexponential law \eqref{eq:GPA} with the characteristic dependence
$-\log S \propto g^2$.  In fact, this argument can justify the
reduction of $D_0$ to an effective diffusion coefficient $D = D_0 (1 +
f(x_0/\ell_{\mathrm d}))$ but does not break the conventional decay
\eqref{eq:GPA}.  Therefore, this explanation fails to rationalize the
localization regime.

From another viewpoint, the argument of ``reduced displacement'' still
relies on the Gaussian phase approximation, in which the signal
attenuation is directly related to the variance of the phase and in
turn to the mean-square displacement of particles.  However, the
distribution of phases is not Gaussian anymore close to a boundary
because of velocity correlations introduced by reflections on the
boundary.

Another flaw in this reasoning is that the relevant scale here is
$\ell_{\mathrm d}$ and not $\ell_g$.  Indeed as shown on
Fig. \ref{fig:correction_Mitra}, the mean-squared displacement is
reduced inside a layer of thickness $\ell_{\mathrm d}$ close to the
boundary (i.e., for particles started between $0$ and $\simeq
4\ell_{\mathrm d}$, see the shaded area).  Naturally, one could argue
that in the regime of $\ell_{\mathrm d} \gg \ell_g$, particles that
travel further than $\ell_g$ would yield a magnetization too small so
that we discard them from the computation of the signal.  This
observation is the basis of the next argument.

\subsection{Competition between confined trajectories and magnetization decay?}
\label{sec:Kiselev}

Let us consider a single impermeable boundary at $x=0$ and introduce a
virtual boundary at $x=\ell$ that particles can freely cross.%
\footnote{
This argument was privately presented to the authors by V.~Kiselev.}
The number of particles $n(\ell)$ that remain confined between the two
boundaries during the whole gradient sequence can be estimated as%
\footnote{
The formula \eqref{eq:n_auxil} is obtained by solving the diffusion
equation inside a slab with an absorbing boundary, which is equivalent
to an interval $(0,\ell)$ with reflections at $0$ and absorptions at
$\ell$.  Precisely, the long-time behavior $n(\ell) \sim (1|u_1)^2
e^{-\lambda_1 t}$ results from the computation of the first eigenmode,
$u_1(x)=\sqrt{2/\ell} \cos(\pi x /(2\ell))$, and the corresponding
eigenvalue $\lambda_1= \pi^2 D_0/(4\ell^2)$, of the diffusion operator
$-D_0 \partial_x^2$.}
%
\begin{equation}  \label{eq:n_auxil}
n(\ell) \sim \ell\exp\left(-\frac{\pi^2 \ell_{\mathrm d}^2}{4 \ell^2}\right)\;,
\end{equation}
and the (non-normalized) signal resulting from those particles follows
from the motional narrowing regime in a slab of width $\ell$ under the
hypothesis $\ell \ll \ell_{\mathrm d}$ \cite{Robertson1966a}:
\begin{equation}  \label{eq:MN_auxil}
s(\ell) = n(\ell) \exp\left(-\frac{\ell^4\ell_{\mathrm d}^2}{120 \ell_g^6}\right)\;.
\end{equation}
Then one evaluates the competition between motional narrowing decay
and ``leakage'' of particles outside the virtual slab by maximizing
the signal with respect to $\ell$.  The maximum is achieved at $\ell
\approx 2.3\, \ell_g$, and the signal becomes
\begin{equation}  \label{eq:s_Kiselev}
s(\ell) \sim \ell_g \exp\left(-0.70\, \ell_{\mathrm d}^2/\ell_g^2\right).
\end{equation}
Although the numerical coefficient $0.70$ is wrong (see below), this
simple reasoning provides the correct form of the signal, i.e., $-\log
s \propto g^{2/3}$.  It shows that the signal is produced by rare
trajectories of particles that stay close to the boundaries of the
domain.  Indeed, the strong diffusion encoding assumption $\ell_g
\ll \ell_{\mathrm d}$ implies that $n(\ell)/\ell$ is very small for
$\ell \sim \ell_g$.

This is an elegant idea that brings additional insights into the
mechanisms behind the localization regime.  However, there are flaws
in this argument, apart from technical issues such as the use of the
motional narrowing formula \eqref{eq:MN_auxil} for non-impermeable
(absorbing) boundaries.

The first one is the use of a virtual perfectly absorbing boundary
that allows for leakage from the slab but prevents the entry of
particles into the slab from the outside.  In that regard, it is hard
to give a physical meaning to $s(\ell)$, since the signal inside the
slab should take into account neighboring particles that enter through
the virtual boundary.  One could argue that the particles from the
outside are discarded because of their strongly attenuated
magnetization.  However, this argument fails for two reasons: (i) if
$\ell \ll \ell_g$, particles that come from a distance $\sim \ell_g$
may enter the virtual slab without experiencing a strong decay and
therefore they cannot be neglected; (ii) if $\ell \gg \ell_g$,
particles from the outside have weak magnetization, but so do
particles inside, and it is not clear why the former might be
neglected with respect to the latter.

However, the most problematic issue is the following: the above
reasoning could be applied exactly the same way to any point of the
medium, regardless of the presence of a boundary.  Instead of
considering a virtual boundary close to the impermeable boundary, one
could consider two virtual boundaries and compute $s(\ell)$ for this
``virtual slab''.  The only change is $n(\ell) \sim \exp\left(-\pi^2
\ell_{\mathrm d}^2/\ell^2\right)$ that in turn yields another
numerical coefficient in Eq. \eqref{eq:s_Kiselev}.  This observation
emphasizes the aforementioned contradictions about the meaning of
$s(\ell)$.

Even though this reasoning yields the correct form of the signal, it
does not explain why the magnetization is localized at the boundary.
In the next section, we suggest a new qualitative interpretation of
the localization regime.  We will see some similarities with the above
discussion that might explain why this wrong reasoning could yield the
correct form of the signal.

\subsection{Symmetry breaking and local effective gradient}
\label{sec:breaking}

Now, we present our own qualitative explanation of the localization
regime.  We will show that the main effect of the boundary is not the
reduction of particles displacements but a {\it symmetry breaking}.
This symmetry breaking produces an effective magnetic field that is
not linear with position but has a V-shape.  Then we show how
localization occurs inside this effective magnetic field.

\begin{figure}[t!] 
\centering
\includegraphics[width=0.99\linewidth]{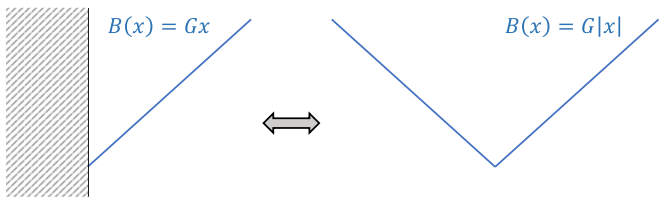} 
\caption{
(left) Impermeable boundary and linear magnetic field. (right) No
boundary and V-shaped magnetic field. Both situations are equivalent
according to the method of images.}
\label{fig:V_potential}
\end{figure}

For simplicity, we consider again a one-dimensional situation, with an
impermeable barrier at $x=0$ and particles diffusing in the half-line
$x > 0$.  The method of images allows one to remove the boundary
provided that each particle on the right half-line is paired with a
``mirror'' particle on the left half-line.  Therefore, the effect of
the impermeable boundary can be taken into account by replacing the
linear magnetic field $B(x)=Gx$ by a V-shape magnetic field
$B(x)=G|x|$, as shown on Fig. \ref{fig:V_potential} (a similar effect
of a parabolic magnetic field was investigated in
\cite{LeDoussal1992a}).  Note that in the Bloch-Torrey equation, the
magnetic field plays the role of an \emph{imaginary} potential, by
analogy with the Schr{\"o}dinger equation.  Although it is tempting to
make a parallel with localization inside a real potential, it is not
evident that the same conclusion would hold for an imaginary
potential.

In order to demonstrate the emergence of the localization phenomenon,
let us write the magnetization in an amplitude-phase representation,
$m(t,x)=A(t,x) e^{i\varphi(t,x)}$, and rewrite the Bloch-Torrey
equation \eqref{eq:BT} in terms of $A$ and $\varphi$:
\begin{subequations}
\begin{align}  \label{eq:BT_A}
&\partial_t A = D_0 A'' -D_0 (\varphi')^2 A \;,\\   \label{eq:BT_varphi}
&\partial_t \varphi = D_0\varphi'' + D_0\frac{A'}{A} \varphi' + G|x|\;,
\end{align}
\end{subequations}
where prime denotes the derivative with respect to $x$.  The initial
conditions are $A(t=0,x)=1$ and $\varphi(t=0,x)=0$.  The first
equation states that $A(t,x)$ obeys a diffusion equation with a
reaction rate $D_0(\varphi')^2$.  The second equation states that
$\varphi$ obeys a diffusion equation with a force term $-D_0 A'/A$ and
a source term $G|x|$.  We emphasize that $\varphi(t,x)$ is a
deterministic function that should not be confused with the random
particle dephasing $\phi$.

\paragraph*{Short times}

\begin{figure}[t!] 
\centering
\includegraphics[width=0.99\linewidth]{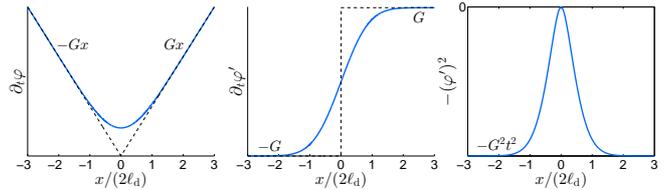} 
\caption{
Plot of $\partial_t \varphi$ (left), $\partial_t \varphi'$ (middle),
and $-(\varphi')^2$ (right) at short times ($\ell_{\mathrm d} \ll
\ell_{g}$).  }
\label{fig:gradient_diff}
\end{figure}

\begin{figure}[t!] 
\centering
\includegraphics[width=0.89\linewidth]{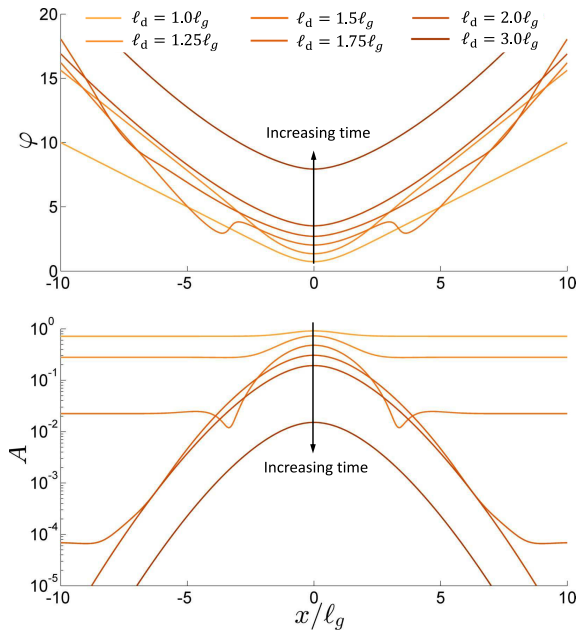} 
\caption{
Time evolution of the magnetization in phase (top) and amplitude
(bottom) representation, for a constant gradient.  The barrier is
located at $x=0$ and the amplitude and phase profiles are reflected
with respect to $x=0$ according to the method of images.  The gradient
length $\ell_g$ is fixed by the gradient, whereas different curves
correspond to six progressively increasing values of $\ell_{\mathrm
d}$ (as indicated in the legend), which are related to time via
$\ell_{\mathrm d} = \sqrt{D_0 t}$.   The magnetization $m(t,x)$
was obtained by solving the Bloch-Torrey equation on a long interval
$(0,L)$ with reflecting endpoints via the matrix formalism
\cite{Grebenkov2007a,Grebenkov2008a}; then its amplitude and phase
were computed.  We also checked that the numerical solution was in
perfect agreement with the exact solution on the half-line derived by
Stoller {\it et al.} \cite{Stoller1991a}.  Refer to the text for
further description.  }
\label{fig:ampli_phase}
\end{figure}

At short times, $A(t,x)$ is nearly constant and the evolution of the
magnetization is dominated by the phase equation
\begin{equation}  \label{eq:varphi_approx}
\partial_t \varphi \simeq D_0 \varphi'' + G|x|\;,
\end{equation}
whose solution is
\begin{align}
\varphi(t,x)&= G \left[ t x\, \mathrm{erf}\biggl(\frac{x}{\sqrt{4D_0 t}}\biggr) - \frac{x^3}{6D_0} 
\biggl(1-\mathrm{erf}\biggl(\frac{x}{\sqrt{4D_0 t}}\biggr)\biggr) \right.\nonumber   \\
& \quad \left.+ \;\frac{4\; \sqrt{D_0 t}}{3\sqrt\pi}  \left(t + \frac{x^2}{4D_0}\right) 
\exp\left(-\frac{x^2}{4D_0 t}\right)\right]. \label{eq:varphi_localization1}
\end{align}
The rate of change of $\varphi$ with time can be interpreted as an
effective magnetic field averaged by diffusion, 
\begin{equation}   \label{eq:varphi_localization2}
\partial_t \varphi(t,x)= G \biggl[x \,\mathrm{erf}\left(\frac{x}{\sqrt{4D_0 t}}\right) 
+ \frac{\sqrt{4D_0t}}{\sqrt\pi} \exp\left(-\frac{x^2}{4D_0t}\right) \biggr]\;,
\end{equation}
and the space derivative of this rate of change is an effective
gradient averaged by diffusion:
\begin{equation}  \label{eq:varphi_localization3}
\partial_t \varphi' (t,x)= G \,\mathrm{erf}\left(\frac{x}{\sqrt{4D_0 t}}\right)\;.
\end{equation}
We have plotted these functions on Fig. \ref{fig:gradient_diff}.  The
main effect of diffusion is to ``smooth'' the V-potential over a
length $\sim \ell_{\mathrm d}$ near $x=0$, resulting in a local
parabolic shape.  In turn, the effective gradient takes smaller values
in this region that translates into smaller values of $(\varphi')^2$.
The results for free diffusion are recovered for $|x| \gg
\ell_{\mathrm d}$, where one gets $\partial_t \varphi' = G$ and
$[\varphi'(t,x)]^2 = (Gt)^2$.  This limits the validity of the
approximate Eq. \eqref{eq:varphi_approx} and its solution in
Eq. \eqref{eq:varphi_localization1} to short times such that $D_0 G^2
t^3/3 \ll 1$, i.e.  $\ell_{\mathrm d}/\ell_g \ll 3^{1/6} \approx 1.2$.
Indeed, these equations rely on the assumption that the amplitude of
the magnetization remains approximately constant in space, i.e. that
the free diffusion decay far from the boundary is not too strong as
compared to the weak decay near $x=0$.  This regime corresponds to
first two curves with $\ell_{\mathrm d}/\ell_g=1.0$ and $1.25$ on
Fig. \ref{fig:ampli_phase}: the amplitude is practically not affected
and the phase profile exhibits the rounded V-shape profile that we
just described.

\paragraph*{Intermediate times}

When the free diffusion decay cannot be neglected anymore, the
evolution of the magnetization enters a second stage of intermediate
times (next two curves with $\ell_{\mathrm d}/\ell_g=1.5$ and $1.75$
on Fig. \ref{fig:ampli_phase}).  The free diffusion decay term $D_0
G^2 t^3$ becomes rapidly very large and the amplitude $A$ decays very
fast, except at the points where $(\varphi')^2$ is significantly
reduced, i.e. in a thin layer of width $\sim\ell_{\mathrm d} \approx
\ell_g$.  As a consequence, the contribution of particles at the sides
(with large phase $\varphi$ and small amplitude $A$) is negligible
compared to that of particles diffusing from the center, which have
small phase $\varphi$ and relatively large amplitude $A$.  Thus, the
phase profile is broadened by diffusion from the center to the sides,
as represented by the force term $(D_0 A'/A) \varphi'$.  In
competition with this broadening effect, the source term $G|x|$ tends
to make the phase profile steeper.  Since the force term enters
through $(D_0 A'/A) \varphi'$, there is a value of $\varphi'$ at which
both effects compensate each other.  In parallel, the evolution of the
amplitude $A$ in Eq. \eqref{eq:BT_A} results from the competition
between diffusion and attenuation.  In fact, the inhomogeneous
attenuation of the amplitude enhances the effect of diffusion, and in
turn diffusion tends to homogenize the amplitude profile.  These
competing effects may even lead to a non-monotonous dependence of the
phase and of the amplitude on $x$ (see the curve with $\ell_{\mathrm
d}/\ell_g=1.5$) but a balance between these two mechanisms is reached
after some time.  Similarly, the temporal evolution of the phase and
the amplitude may be non-monotonous (see, e.g., the curves on
Fig. \ref{fig:ampli_phase} at $x/\ell_{\rm g} \approx 3.5$).

\paragraph*{Long times}   
   
In the final stage, a dynamic balance between two competing effects is
set (two last curves with $\ell_{\mathrm d}/\ell_g=2.0$ and $3.0$ on
Fig. \ref{fig:ampli_phase}).  Diffusion tends to broaden the amplitude
profile, but the strong decay $-(\varphi')^2$ destroys the
magnetization outside of the region $|x| \lesssim \ell_g$.  Therefore,
the situation is analogous to that of a slab of width $\sim \ell_g$
with absorbing boundaries, hence the decay $-\log A \sim \ell_{\mathrm
d}^2/\ell_g^2$.  The source term $G|x|$ tends to make the phase
profile steeper but the force term $-D_0 A'/A$ broadens it by
``pushing'' towards high $|x|$.  In other words, particles at the
center with a (relatively) strong magnetization diffuse away from the
center and outweigh the contribution of particles at the sides that
have a much weaker magnetization.  Therefore, the source term $G|x|$
contributes only up to $|x|\approx \ell_g$, and the phase profile
translates upwards as $\sim G \ell_g t= \ell_{\mathrm d}^2/\ell_g^2$.
These conclusions reproduce exactly the behavior of the magnetization
in the localization regime.
 
In analogy to the argument of Sec. \ref{sec:Kiselev}, we obtain that
the localization phenomenon at long times is similar to diffusion
inside a slab of width $\ell_g$ with absorbing boundaries.  In other
words, the signal in the localization regime is produced by rare
trajectories that remain close to the boundary at all times.  However,
we do not rely on the motional narrowing regime formula and the effect
of the boundary is explicitly taken into account as a symmetry
breaking of the phase profile: the phase profile becomes even instead
of being odd that leads to a region with reduced decay rate
$(\varphi')^2$.  In particular, in the absence of a boundary, the
magnetic field profile is linear and there is no region of space with
a reduced effective gradient, thus no localization.
 
In the next paragraph, we further emphasize the mechanism of the
localization regime by taking into account the size $\ell_{\mathrm s}$
of the domain and by investigating qualitatively the transition
between the motional narrowing regime and the localization regime.

\subsection{Localization regime and motional narrowing regime}

\begin{figure}[t!] 
\centering
\includegraphics[width=0.99\linewidth]{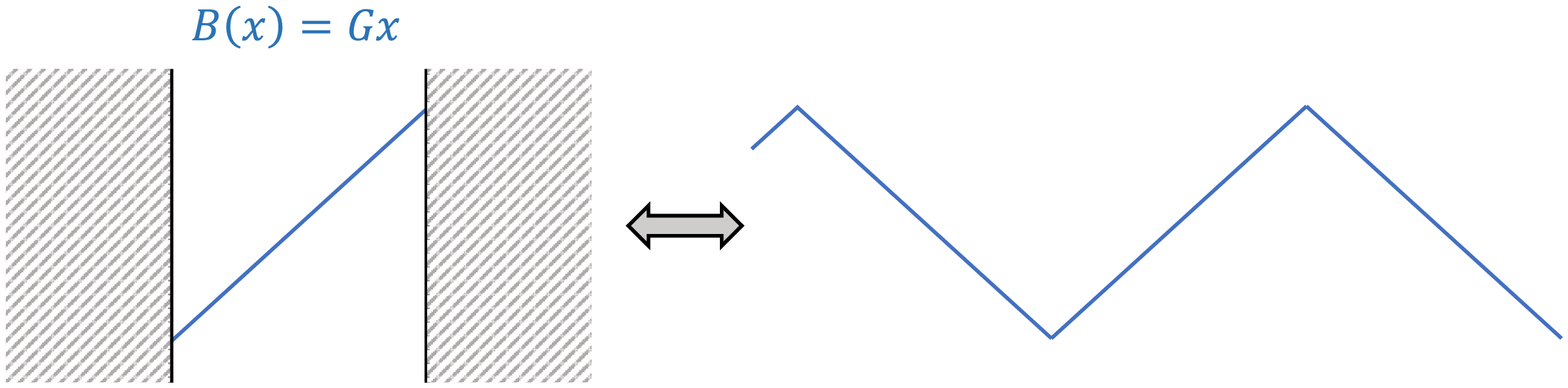} 
\caption{
(left) Slab with impermeable boundaries and linear magnetic
field. (right) No boundary and periodic triangular profile.  Both
situations are equivalent according to the method of images.}
\label{fig:V_potential_period}
\end{figure}

\begin{figure}[t!] 
\centering
\includegraphics[width=0.99\linewidth]{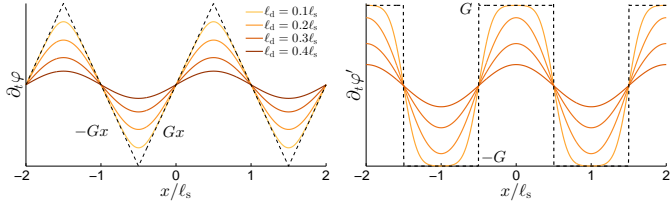} 
\caption{
Effective magnetic field (left) and effective gradient (right)
averaged by diffusion for $\ell_{\mathrm d} \ll \ell_g$, at various
ratios $\ell_{\mathrm d}/\ell_{\mathrm s}$. As time increases, both
functions become rounder but also weaker because of compensation
between positive and negative parts. Note that the slab corresponds to
$-1/2 \leq x/\ell_{\mathrm s} \leq 1/2$ and is repeated periodically
according to the method of images.  }
\label{fig:gradient_diff_period}
\end{figure}

We employ the same qualitative description as above, but now we
consider a finite slab of width $\ell_{\mathrm s}$.  As illustrated on
Fig. \ref{fig:V_potential_period}, the method of images yields a
periodic triangular magnetic field, with period $2\ell_{\mathrm s}$.
To obtain the phase profile $\varphi$ at short times ($\ell_{\mathrm
d} \ll \ell_g$), we solve the diffusion equation (\ref{eq:BT_varphi})
without the force term $(D_0 A'/A)\varphi'$ and get
\begin{align} \nonumber
\varphi(t,x) & \simeq \frac{\ell_{\mathrm s}^3}{\ell_g^3} \sum_{n=0}^\infty \frac{(-1)^n}{4\pi^4 (n+1/2)^4} 
\sin\left((2n+1)\pi {x}/{\ell_{\mathrm s}}\right) \\
& \times \left[1-e^{-(2n+1)^2 \pi^2 \ell_{\mathrm d}^2/\ell_{\mathrm s}^2}\right]\;.
\label{eq:triangle_diffusion}
\end{align}
The effective magnetic field $\partial_t \varphi$ and the effective
gradient $\partial_t \varphi'$ averaged by diffusion follow
immediately from this solution and are plotted on
Fig. \ref{fig:gradient_diff_period}.  One can see that as time
increases, the effective field and gradient become rounder and weaker
because of compensation between positive and negative parts.
Depending on the range of validity of the above formulas, one is
naturally led to distinguish between two regimes.  While the
transition between these two regimes was rigorously analyzed in
\cite{Grebenkov2014b} by using the exact solution for the
magnetization in terms of Airy functions, we provide here much simpler
qualitative arguments.
 
\paragraph*{Localization regime (large slab, strong gradient)} 

Let us first consider the case $\ell_g \ll \ell_{\mathrm s}$
corresponding to the localization regime.  For example, if $\ell_g =
0.1 \ell_{\mathrm s}$, the above formula for $\varphi$ (and its
consequences for $\partial_t \varphi$ and $\partial_t \varphi'$) are
valid until $\ell_{\mathrm d} \lesssim 0.1 \ell_{\mathrm s}$ (see
light yellow curves on Fig. \ref{fig:gradient_diff_period}).  Since
$\ell_{\mathrm d} \ll \ell_{\mathrm s}$, the effective magnetic field
profile is close to a triangular shape with small parabolic parts,
similarly to the left panel of Fig. \ref{fig:gradient_diff}.  The
discussion of Sec. \ref{sec:breaking} applies without any modification
to this regime, and the magnetization is localized at the boundaries
of the slab.

\paragraph*{Motional narrowing regime (small slab, weak gradient)}

The opposite case $\ell_{\mathrm s} \ll\ell_g$ corresponds to the
motional narrowing regime.  In that case, the above formulas for
$\partial_t \varphi$, $\partial_t \varphi'$ are valid over a very long
time range (corresponding to the dark brown curves on
Fig. \ref{fig:gradient_diff_period}).  In particular, for
$\ell_{\mathrm s} \lesssim \ell_{\mathrm d} \ll \ell_g$, the phase
profile reaches the steady-state form, which is obtained from
Eq. \eqref{eq:triangle_diffusion}:
\begin{align}  \nonumber
\varphi(t,x)& \xrightarrow[t \to \infty]{}
\frac{\ell_{\mathrm s}^3}{\ell_g^3} \sum_{n=0}^\infty 
\frac{(-1)^n}{4\pi^4 (n+1/2)^4} \sin\big((2n+1)\pi {x}/{\ell_{\mathrm s}}\big)  \\
&=\frac{\ell_{\mathrm s}^3}{24 \ell_g^3} \frac{x}{\ell_{\mathrm s}}
\left(3-4 \frac{x^2}{\ell_{\mathrm s}^2}\right)  \quad \bigl(-\tfrac{1}{2} \leq x/\ell_{\mathrm s} \leq \tfrac{1}{2}\bigr)\;.
\label{eq:varphi_approx2}
\end{align}
This expression gives immediately the decay rate,
\begin{equation}
D_0 [\varphi']^2 \approx\frac{1}{64}\frac{D_0 \ell_{\mathrm s}^4}{\ell_g^6} 
\left[ 1 - \left(\frac{2x}{\ell_{\mathrm s}}\right)^2 \right]^2 \;,
\end{equation}
while its average over the half-period,
\begin{equation}
\frac{1}{\ell_{\mathrm s}} \int_{-\ell_{\mathrm s}/2}^{\ell_{\mathrm s}/2} D_0 [\varphi'(t,x)]^2 \,\mathrm dx 
\approx \frac{D_0\ell_{\mathrm s}^4}{120 \ell_g^6} \;,
\end{equation}
yields an average signal decay 
\begin{equation}
S \approx \exp\left(- \frac{1}{120} \ell_{\mathrm d}^2 \ell_{\mathrm s}^4/\ell_g^6 \right), 
\end{equation}
which is the exact result for the motional narrowing regime
\cite{Robertson1966a,Neuman1974a,Grebenkov2007a}.  

At long times, the decay of the signal becomes significant, and one
may wonder about the validity of the above result.  Actually, one can
see that the decay of the signal occurs on a time scale $\ell_g^6/(D_0
\ell_{\mathrm s}^4)$ much larger than the diffusion time scale
$\ell_{\mathrm s}^2/D_0$.  As a consequence the diffusion term
Eq. \eqref{eq:BT_A} flattens any inhomogeneity in the amplitude
profile.  In turn, since the amplitude profile is nearly homogeneous
at all times, the formula \eqref{eq:varphi_approx2} for $\varphi$,
that relied on neglecting the force term $D_0 A'/A$, is always valid
at long times.

\subsection{Breakdown of the Gaussian phase approximation}

The motional narrowing regime may be obtained from the central limit
theorem applied to successive explorations of a bounded domain
\cite{Neuman1974a}.  The main hypothesis behind this reasoning is that
any particle ``loses memory'' of its initial position after a time
$\sim \ell_{\mathrm s}^2/D_0$.  This hypothesis allows one to treat
the accumulated phases over successive explorations of the domain as
independent from each other, that is a crucial assumption of the
central limit theorem.  This argument implies that the distribution of
the random phase $\phi$ is Gaussian, whereas the signal, which can be
understood as a characteristic function $\phi$, is a Gaussian function
of the gradient strength: $-\log S \sim G^2 \ell_{\mathrm s}^4 t /
D_0$.  Since this reasoning relies on the central limit theorem, it
seems very robust and it is \textit{a priori} not clear why it would
break down if the gradient length is much smaller than the pore
diameter (i.e., $\ell_g \ll \ell_{\mathrm s}$).

The above computation reveals that in the localization regime, a small
fraction of particles, of order of $\exp(-\ell_{\mathrm
d}^2/\ell_g^2)$, remains close to the boundary and dominates the
signal due to the local symmetry breaking caused by the boundary.  In
terms of accumulated phase, this means that the velocity correlations
introduced by the boundary make the phase distribution non-Gaussian
for particles close to the boundary, and its contribution dominates in
the signal at high gradients.  It is worth noting, however, that the
regime $\ell_g \ll \ell_{\mathrm s} \ll \ell_{\mathrm d}$ would yield
a very weak signal anyway.

\section{Discussion}

\subsection{Universal character of the localization regime}

The simple explicit form \eqref{eq:S_free} of the signal for free
diffusion is the first result that a student learns about diffusion
NMR.  Its natural extension into Eq. \eqref{eq:GPA} via the Gaussian
phase approximation, which is always valid at small gradients, and
numerous experimental observations re-enforce the common belief that
the monoexponential form \eqref{eq:GPA} is the good starting point to
analyze signals at higher gradients.  On one hand, there are various
models based on Eq. \eqref{eq:GPA}; on the other hand, the cumulant
expansion aims at improving the Gaussian phase approximation by adding
perturbative corrections.  In this light, the unusual,
$G^{2/3}$-behavior of the signal in the localization regime looks
indeed as a pathology, far apart from the common trend.  In this
section, we briefly describe the major flaws of this dogmatic view and
emphasize on the universal character of the localization regime.

The non-perturbative approach to diffusion NMR, initiated by Stoller
{\it et al.} \cite{Stoller1991a} and then developed in a series of
publications
\cite{deSwiet1994a,Hurlimann1995a,Grebenkov2014b,Herberthson2017a,Grebenkov2017a,Grebenkov2018a,Grebenkov2018b,Almog2018a,Almog2019a,Moutal2019b,Moutal2020a},
relies on the spectral analysis of the Bloch-Torrey operator,
${\mathcal B}_g = - \nabla^2 + i \ell_g^{-3} x$, where $x$ is the
coordinate along the gradient direction.  For a pulsed-gradient
spin-echo sequence shown on Fig. \ref{fig:PGSE}, this operator governs
the time evolution of the transverse magnetization.  If the spectrum
of ${\mathcal B}_g$ is {\it discrete}, the signal can be represented as
a spectral expansion:
\begin{equation}  \label{eq:S_spectral}
S = \sum\limits_{j,k} c_{j,k}(\ell_g,\Delta-\delta) \, \exp\bigl(-(\lambda_j(\ell_g) + \lambda_k^*(\ell_g))D_0 \delta\bigr) ,
\end{equation}
where $\lambda_j(\ell_g)$ are the eigenvalues of ${\mathcal B}_g$, and
$c_{j,k}(\ell_g,\Delta-\delta)$ are the coefficients based on its
eigenfunctions \cite{deSwiet1994a,Grebenkov2014b}.  Since the
Bloch-Torrey operator is not Hermitian, the eigenvalues are in general
complex-valued: their real part determines the decay of the signal
whereas the imaginary part is responsible for eventual oscillations.
At high gradients, the eigenvalues exhibit a universal scaling
behavior:
\begin{equation} 
\mathrm{Re}(\lambda_j(\ell_g)) \propto \ell_g^{-2} 
\biggl(1 + O\bigl((\ell_{\rm s}/\ell_g)^{\frac{1}{2}}\bigr)\biggr)  \quad \textrm{as}~ \ell_g \to 0.
\end{equation}
For extended gradient pulses (when $\delta$ is large enough), the term
containing the eigenvalue $\lambda_1$ with the smallest real part
provides the major contribution to the signal:
\begin{align}  \nonumber
S & \approx c_{1,1}(\ell_g,\Delta - \delta) \, \exp\bigl(-2 \mathrm{Re}(\lambda_1(\ell_g)) D_0 \delta\bigr) \\  \label{eq:Sloc2}
& \sim \exp\bigl(-|a'_1| \ell_{\mathrm{d}}^2/\ell_g^2\bigr) = \exp\bigl(-2|a'_1| D_0^{1/3} G^{2/3} \delta\bigr),
\end{align}
where $|a'_1| \approx 1.0188$ is a universal numerical prefactor.  As
we discussed earlier, this behavior is tightly related to the
localization of the associated eigenfunctions near the boundary
points, at which the gradient is perpendicular to that boundary.  The
universality of Eq. \eqref{eq:Sloc2} follows from the local character
of this localization phenomenon: when the gradient length $\ell_g$ is
the smallest scale of the problem, the boundary looks flat on this
scale.  The effect of other terms in Eq. \eqref{eq:S_spectral} and the
next-order corrections to the eigenvalues are discussed in
\cite{Moutal2019b}.

As one can see, the above $G^{2/3}$-behavior of the signal is very
general, whenever the spectrum of the Bloch-Torrey operator is
discrete.  This is true for any bounded domain, i.e., when the nuclei
diffuse within a finite-size sample.  Unexpectedly, the recent works
have shown that the spectrum is also discrete for a large class of
unbounded and periodic domains
\cite{Almog2018a,Almog2019a,Moutal2019b,Moutal2020a}.  This is a
counter-intuitive result because the spectrum of the Laplace operator,
${\mathcal B}_0 = -\nabla^2$, which is obtained in the limit of no
gradient, is continuous in these domains.  In other words, the limit
$g\to 0$, in which the discrete spectrum of the Bloch-Torrey operator
${\mathcal B}_g$ transforms into the continuous spectrum of ${\mathcal
B}_0$, is singular.  This general mathematical argument implies that a
perturbative construction of the signal that starts from the purely
diffusive operator ${\mathcal B}_0$ and treats the extended gradient
pulse as a perturbation, is doomed to fail in such unbounded or
periodic domains.  Even though this discussion may sound rather
abstract, it reveals some fundamental mathematical flaws in commonly
used perturbative theories%
\footnote{
We hasten to emphasize that we speak here about perturbative theories
(such as a cumulant expansion), in which the gradient term is treated
as a perturbation to the diffusion operator.  This perturbative
approach should not be confused with the effective medium theory,
which was developed for infinitely narrow gradient pulses
\cite{Novikov2010a}.  In the latter case, the original non-Hermitian
problem is reduced to a purely diffusive problem in a heterogeneous
medium, and the effect of heterogeneities is treated perturbatively.
Here, one deals with the Hermitian diffusion operator, for which there
is no localization phenomenon, and the perturbative approach is
valid.}
of diffusion NMR and urges for new developments in this field.

Even for bounded domains, the non-Hermitian character of the
Bloch-Torrey operator makes the study of diffusion NMR challenging but
very rich.  In particular, the spectrum of the Bloch-Torrey operator
in bounded domains typically has branching or bifurcation points,
i.e., there are critical values of the gradient $G$, at which two
eigenvalues coalesce.  This effect was first discussed by Stoller {\it
et al.} \cite{Stoller1991a} for an interval and later found in other
domains \cite{Moutal2019b,Moutal2020a}.  From the mathematical point
of view, such branching points induce non-analyticity in the spectrum
and potentially in other related quantities such as the magnetization
and the signal.  In particular, the cumulant expansion may diverge
beyond the smallest critical gradient that would be the ultimate range
of validity of a perturbative approach.  Further understanding of the
limitations of the cumulant expansion presents an interesting
perspective for future research.

Anticipating further mathematical progress in this direction, we
conjectured that the spectrum of the Bloch-Torrey operator is discrete
for any nontrivial domain \cite{Grebenkov2018a}.  If this conjecture
will be proved, the general spectral representation
\eqref{eq:S_spectral} and the consequent localization regime at high
extended gradients will become a common rule, whereas the free
diffusion signal \eqref{eq:S_free} will be an exception from that
rule.  However, we hasten to stress that such a paradigm shift in the
theoretical description of diffusion NMR does not diminish the
importance of the Gaussian phase approximation and the related models
developed at small gradients.  On the contrary, bridging the
small-gradient and high-gradient asymptotic forms of the signal
presents one of the major theoretical challenges for future research.

\subsection{Summary of diffusion NMR regimes}
\label{section:extended_gradient}

\begin{figure}[t!] 
\centering
\includegraphics[width=0.99\linewidth]{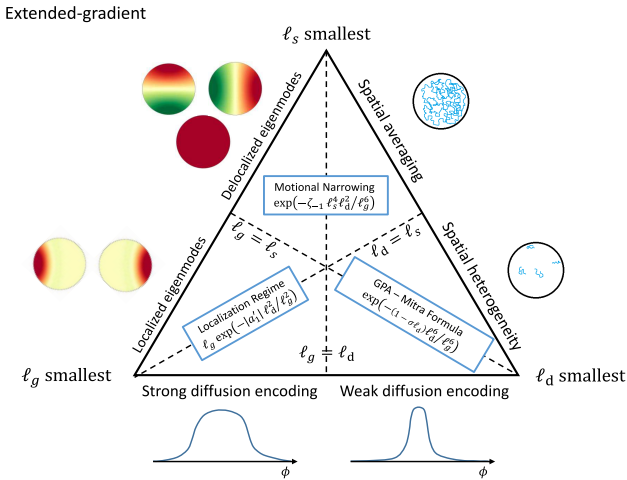} 
\caption{
Schematic representation of three regimes of diffusion NMR for
extended-gradient pulse experiments.  Three length scales
$\ell_{\mathrm d}$, $\ell_{\mathrm s}$ and $\ell_g$ are defined in the
text.  In addition, $\sigma$ is related to the surface-to-volume ratio
of the medium \cite{Mitra1992a,Mitra1993a,Moutal2019a}, $\zeta_{-1}$
is a shape-dependent coefficient \cite{Grebenkov2007a}, and $|a'_1|
\approx 1.0188$ is a universal numerical prefactor
\cite{Stoller1991a}.  Colored patterns inside the disks on the left
illustrate the localized and delocalized eigenmodes of the
Bloch-Torrey operator governing the evolution of the transverse
magnetization \cite{Moutal2019b,Moutal2020a}. }
\label{fig:schema_these_extended}
\end{figure}

The above discussion clarified the opposite effects of diffusion and
gradient encoding onto the magnetization.  The competition of these
two effects, characterized respectively by the length scales
$\ell_{\mathrm d}$ and $\ell_g$, produces a variety of diffusion NMR
regimes \cite{Axelrod2001a,Grebenkov2007a}.  H\"urlimann {\it et al.}
identified three major regimes by plotting a schematic diagram 
in the parameters' plane $(\ell_g/\ell_{\mathrm s} , \ell_{\mathrm
d}/\ell_{\mathrm s})$ \cite{Hurlimann1995a} (see also
\cite{Grebenkov2007a}).  As each of these regimes emerges when one of
the three length scales ($\ell_{\mathrm d}$, $\ell_{\mathrm s}$ and
$\ell_g$) is the smallest, we find convenient to summarize these
regimes by a triangle shown on Fig. \ref{fig:schema_these_extended}:

(i) When the diffusion length $\ell_{\mathrm d}$ is the smallest, the
nuclei travel short distances and acquire weak dephasing.  This is the
slow-diffusion or short-time regime, in which the effective diffusion
coefficient $D$ of the nuclei is reduced by the microstructure.  The
signal attenuation exhibits approximately the monoexponential form
(\ref{eq:GPA}), whereas some geometric characteristics of the medium
such as its surface-to-volume ratio can be estimated
\cite{Mitra1992a,Mitra1993a,Moutal2019a}.

(ii) When the structural length $\ell_{\mathrm s}$ is the smallest,
the nuclei explore the confining domain and average the magnetic field
inhomogeneities.  In this motional-narrowing or long-time regime, the
Gaussian phase approximation is again valid, and the signal formally
admits the monoexponential form \cite{Robertson1966a,Neuman1974a}.
However, the apparent diffusion coefficient $D$ is inversely
proportional to $D_0$ and strongly depends on the echo time and the
confinement scale $\ell_{\mathrm s}$.

(iii) When the gradient length $\ell_g$ is the smallest, the
transverse magnetization is localized near the boundary, the Gaussian
phase approximation fails, and the signal exhibits ``anomalous''
scaling \eqref{eq:Sloc}.

While the triangular diagram on Fig. \ref{fig:schema_these_extended}
gives a panora\-ma of diffusion NMR regimes, it is still a schematic
simplification of the complexity and variety of this phenomenon.  In
fact, most biological or mineral samples exhibit multiple length
scales that so different regimes may co-exist or emerge progressively.
In addition, surface relaxation $\rho$ or membrane permeability
$\kappa$ would further complicate this picture by reducing the
magnetization near the boundary either by relaxation or diffusive
exchange.  Both effects induce their own length scale, either
$\ell_{\rho} = D_0/\rho$ or $\ell_{\kappa} = D_0/\kappa$, which should
be included into the analysis, transforming the triangle diagram into
a tetrahedron with four length scales.  In this way, one can
incorporate the signal decay due to the surface relaxation or
permeation without gradient encoding
\cite{Brownstein1979a} but also investigate their coupling
\cite{Grebenkov2014b}.  Finally, we focused on the most basic gradient
sequence with two equal rectangular pulses, whereas the temporal
profile of the gradient sequence can be easily manipulated in
experiments.  For instance, the application of unequal rectangular
pulses led to pore imaging modality \cite{Kuder2013a,Hertel2013a},
while isotropic diffusion weighting and more elaborated encoding
schemes brought novel insights onto anisotropy studies
\cite{Eriksson2013a,Topgaard2017a,Moutal2019a}.

\subsection{Localization versus diffusion-diffraction regimes}
\label{sec:narrow}

\begin{figure}[t!] 
\centering
\includegraphics[width=0.99\linewidth]{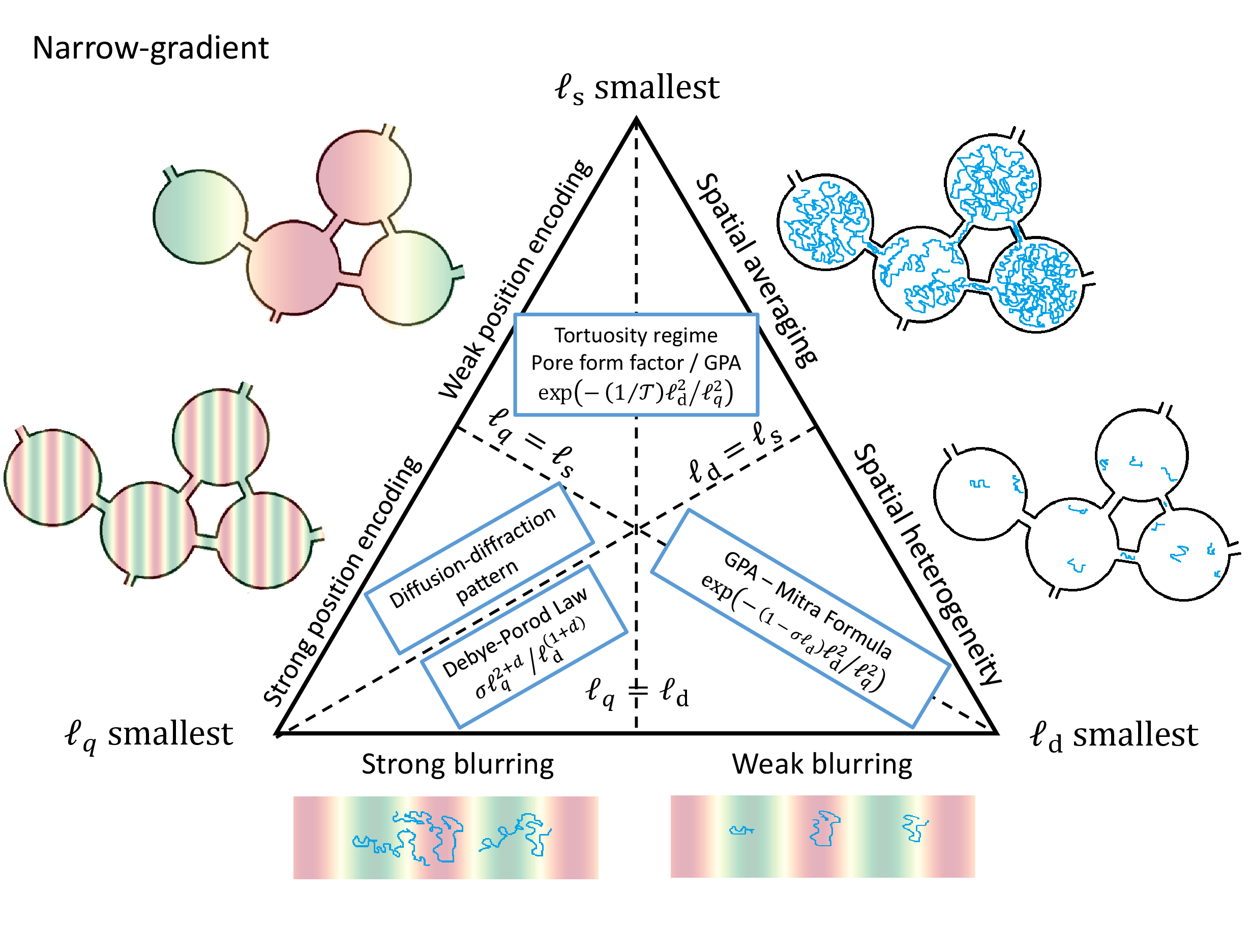} 
\caption{
Schematic representation of different regimes of diffusion NMR for
narrow-gradient pulse experiments.  Three length scales $\ell_{\mathrm
d}$, $\ell_{\mathrm s}$ and $\ell_q$ are defined in the text.  In
addition, $\sigma$ is related to the surface-to-volume ratio of the
medium \cite{Mitra1992a,Mitra1993a,Moutal2019a}, ${\mathcal T} =
D_0/D(t=\infty)$ is the tortuosity of the medium \cite{Latour1995a},
and $d$ is the space dimension.  Colored patterns on the left
illustrate the phase patterns induced by a narrow-gradient pulse.}
\label{fig:schema_these_narrow}
\end{figure}

High gradient pulses have been employed in diffusion NMR for many
decades but they are usually associated with the so-called
narrow-pulse approximation \cite{Stejskal1965b,Tanner1968a}.  Let us
consider again the PGSE sequence shown on Fig. \ref{fig:PGSE}.  If the
gradient duration $\delta$ is very short so that diffusion of the
nuclei over time $\delta$ can be neglected, then the effect of a
single narrow-gradient pulse is simply to multiply the magnetization
by the phase pattern $e^{-i(\mathbf{q} \cdot \mathbf{r})}$ with no
attenuation, where $\mathbf{q} = \delta \mathbf{G}$ is the associated
wavevector.  In turn, the attenuation of the magnetization is caused
by the subsequent diffusion step of duration $\Delta$ that ``blurs''
the phase pattern of period $\ell_q = 1/|\mathbf{q}|$ (up to a factor
$2\pi$).  Given that $\delta$ is short, one needs to apply high
gradients to reduce $\ell_q$ in order to probe the microstructure with
comparable length scales.  Moreover, the theoretical description
relies on the limit of infinitely narrow ($\delta \to 0$) but
infinitely strong ($|\mathbf{G}|\to \infty$) gradient pulses with
fixed $\mathbf{q}$.  The competition between $\ell_q$, $\ell_{\mathrm
d}$ (here, $\ell_{\mathrm d} = \sqrt{D_0 \Delta}$) and the confining
length $\ell_{\mathrm s}$ yields three major regimes summarized on
Fig. \ref{fig:schema_these_narrow}:

(i) When the diffusion length $\ell_{\mathrm d}$ is the smallest, one
retrieves the same slow-diffusion regime as with extended gradient
pulses in Sec. \ref{section:extended_gradient}.

(ii) When the structural length $\ell_{\mathrm s}$ is the smallest,
the nuclei explore the confining domain and can thus probe its global
structure such as connectivity, tortuosity and disorder
\cite{Latour1995a,Novikov2011a,Novikov2014a}.

(iii) When the phase pattern period $\ell_q$ is the smallest, the
blurring of the phase pattern by diffusion renders the signal very
sensitive to the microstructure.  In particular, the signal exhibits a
power law decay (Debye-Porod law)
\cite{Debye1957a,Sen1995a,Froehlich2006a}, possibly with oscillations
(known as diffusion-diffraction patterns), from which some structural
properties of the medium can be determined
\cite{Callaghan1991b,Callaghan1992a,Coy1994a,Callaghan1995a,Linse1995a}.

As for extended-gradient pulses with the smallest $\ell_g$, the
diffusion-diffraction regime with the smallest $\ell_q$ results from
the strong coupling between diffusion and gradient encoding and thus
potentially allows one to infer fine structural properties.  However,
such a high sensitivity to the microstructure can also be considered
as a drawback: an unavailable variability of shapes and sizes in the
microstructure leads to superposition of diffusion-diffraction
patterns and may partially or fully destroy them.  In this regard, the
local behavior of the magnetization for extended-gradient pulses can
make the localization regime more robust against such averages,
keeping its sensitivity.

Even though both the localization and diffusion-diffrac\-tion regimes
emerge at high gradients, there are several fundamental differences
between them.  The signal attenuation occurs due to coupled effects of
diffusion and encoding during the gradient pulse in the former case,
whereas it is purely blurring effect of diffusion in-between two
gradient pulses in the latter case (in particular, if the inter-pulse
time $\Delta$ is set to $\delta \approx 0$, there is no signal
attenuation).  This effect can also be quantified by looking at the
$b$-value that characterizes the overall gradient encoding by the
pulsed-gradient sequence.  Setting $\Delta = \delta$, one gets $b =
\tfrac{2}{3} G^2 \delta^3 = \tfrac{2}{3} q^2 \delta \to 0$ in the
narrow-pulse limit $\delta \to 0$.  In contrast, for extended-gradient
pulses with fixed $\delta$, the localization regime corresponds to $G
\to \infty$ so that the $b$-value also grows to infinity.

\subsection{A practical guideline}

To date, there is no experimental protocol exploiting the advantages
of the localization regime, notably, its high sensitivity to the
microstructure.  On one hand, such experiments have to be realized
with relatively high signal-to-noise (SNR) ratios, certainly above 10
and better up to 1000.  Indeed, as illustrated on Fig. \ref{fig:magn}
(see also related figures in \cite{Grebenkov2014b,Moutal2019b}),
deviations from the monoexponential behavior become notable when $S
\lesssim 0.1$.  In other words, ``large'' signals decay in a similar
(monoexponential) way, whereas the behavior of ``small'' signals is
more specific and thus more sensitive to structural changes.
On the other hand, further theoretical progress is needed to handle
heterogeneities of the sample such as variability in shapes and sides.
While the leading asymptotic term \eqref{eq:Sloc2} is universal, it is
usually not sufficient to accurately describe the signal
\cite{Moutal2019b}, and one has to resort to the general spectral
expansion \eqref{eq:S_spectral} and to investigate how the spectral
properties of the Bloch-Torrey operator are related to the
microstructure.  An extension to a general temporal profile of the
gradient (beyond the considered rectangular gradient pulses) is
another important problem for future research.

Even though the advantages of the localization regime are not yet
exploitable, the partial localization of the magnetization near the
boundaries still affects the signal and may lead to deviations from
its monoexponential decay.
However, such deviations can also be caused by superposition of
signals from isolated pores of variable sizes and shapes, exchange
between multiple compartments, surface relaxation or permeation,
susceptibility-induced gradients, and other mechanisms.
Quite often, non-mono\-exponential signals are analyzed by fitting to
some model formulas, whose adjustable parameters are then related to
the microstructure or used as biomarkers.  For instance, the
bi-exponential model was used to fit the spin-echo signal acquired in
the brain, and the estimated fraction of the intracellular space was
suggested as a biomarker of cell swelling after an ischemic insult
\cite{Niendorf1996a}.  However, it is important to stress that an
accurate fit does not necessarily justify the model underlying the
fitting formula \cite{Grebenkov2010a}.  In particular, the
bi-exponential model accurately fits the signal shown on
Fig. \ref{fig:magn}(c) for moderate $b$-values but any microstructural
interpretation of its parameters is meaningless for this setting.  As
the origin of the non-monoexponential decay can hardly be identified
from a single measurement, it is recommended to conduct a series of
experiments by varying the gradient profile, e.g., the amplitude of
the gradient pulse, its duration, and the inter-pulse time.  An
exploration towards high gradients can also be beneficial.  When such
a systematic investigation is not feasible (e.g., in medical imaging),
one can at least estimate the typical length scales $\ell_{\mathrm
d}$, $\ell_g$ (or $\ell_q$) and $\ell_{\mathrm s}$ in order to
position the experimental setting on the diagrams shown on
Figs. \ref{fig:schema_these_extended} and
\ref{fig:schema_these_narrow}.  Numerical simulations on model
structures representative of the studied sample can bring
complementary insights.

\section{Conclusion}

In this paper, we argued about the fundamental importance of the
localization regime, which was generally ignored in diffusion NMR.  We
provided the first qualitative, physically-appealing description of
the localization mechanism, beyond a formal mathematical solution of
the Bloch-Torrey equation.  Contrarily to former hand-waving
arguments, we identified the symmetry breaking of the gradient profile
by a reflecting boundary as the origin of localization.  We emphasized
the relation between the ``anomalous'', $G^{2/3}$-behavior of the
signal, the localization of the transverse magnetization, and the
discrete spectrum of the governing Bloch-Torrey operator.  Recent
mathematical advances in the spectral analysis of this operator
support our claim about the universal character of the localization
regimes at high extended gradient pulses.  In this light, the
localization regime is not a pathologic exception but a common rule.
Moreover, we discussed fundamental limitations of the current
perturbative approaches when the gradient increases.  This analysis
suggests that the current paradigm in diffusion NMR needs to
re-considered towards the development of non-perturbative methods.
Quite surprisingly, only now, after seventy years of intensive
research in this field, we start to realize how partial is our
understanding on the signal attenuation due to the coupled effects of
diffusion and magnetic field.  In particular, revealing limitations of
the cumulant expansion due to branching points of the spectrum and
mastering a transition between small-gradient and high-gradient
asymptotic regimes remain open problems.

At the same time, the localization regime and related questions are
not purely theoretical.  An experimental study with a hyperpolarized
gas has recently shown the emergence of the localization regime inside
cylindrical pores and outside an array of cylindrical obstacles
\cite{Moutal2019b}.  This study was realized on a clinical MRI scanner
and showed severe deviations from the Gaussian phase approximation at
gradients as small as 7~mT/m (see also Fig. \ref{fig:magn}).  As a
consequence, most diffusion NMR studies nowadays would show such
deviations and {\it may} be affected by the localization regime.  Even
though the localization regime is clearly not the only cause of such
deviations, its ignorance may result in false and misleading
interpretations of experimental results
\cite{Grebenkov2010a,Grebenkov2018a}.  More generally, the high
sensitivity of the signal to the microstructure in the localization
regime presents an unexplored opportunity for new imaging modalities.
While the weak signal does not still allow one to exploit this
opportunity today, further studies of the strong coupling between
diffusion and magnetic field encoding will hopefully overcome this
practical limitation.

\section*{Acknowledgments}
D.S.G. acknowledges a partial financial support from the Alexander von
Humboldt Foundation through a Bessel Research Award.

\appendix
\renewcommand{\thefigure}{\arabic{figure}}

\section{Two length scales associated to the gradient}
\label{section:scales}

In this Appendix, we elaborate our qualitative discussion about two
fundamental length scales associated to the gradient: the gradient
length $\ell_g$ and the phase pattern period $\ell_q$.  These two
length scales have different physical interpretation and are somewhat
``exclusive'': while $\ell_g$ is better suited to discuss the behavior
of extended-gradient pulse experiments, $\ell_q$ is better suited to
the opposite case of narrow-gradient pulse experiments.

\subsection*{Gradient length $\ell_g$}
\label{section:gradient_length}

\begin{figure}[t!]
\centering
\includegraphics[width=100mm]{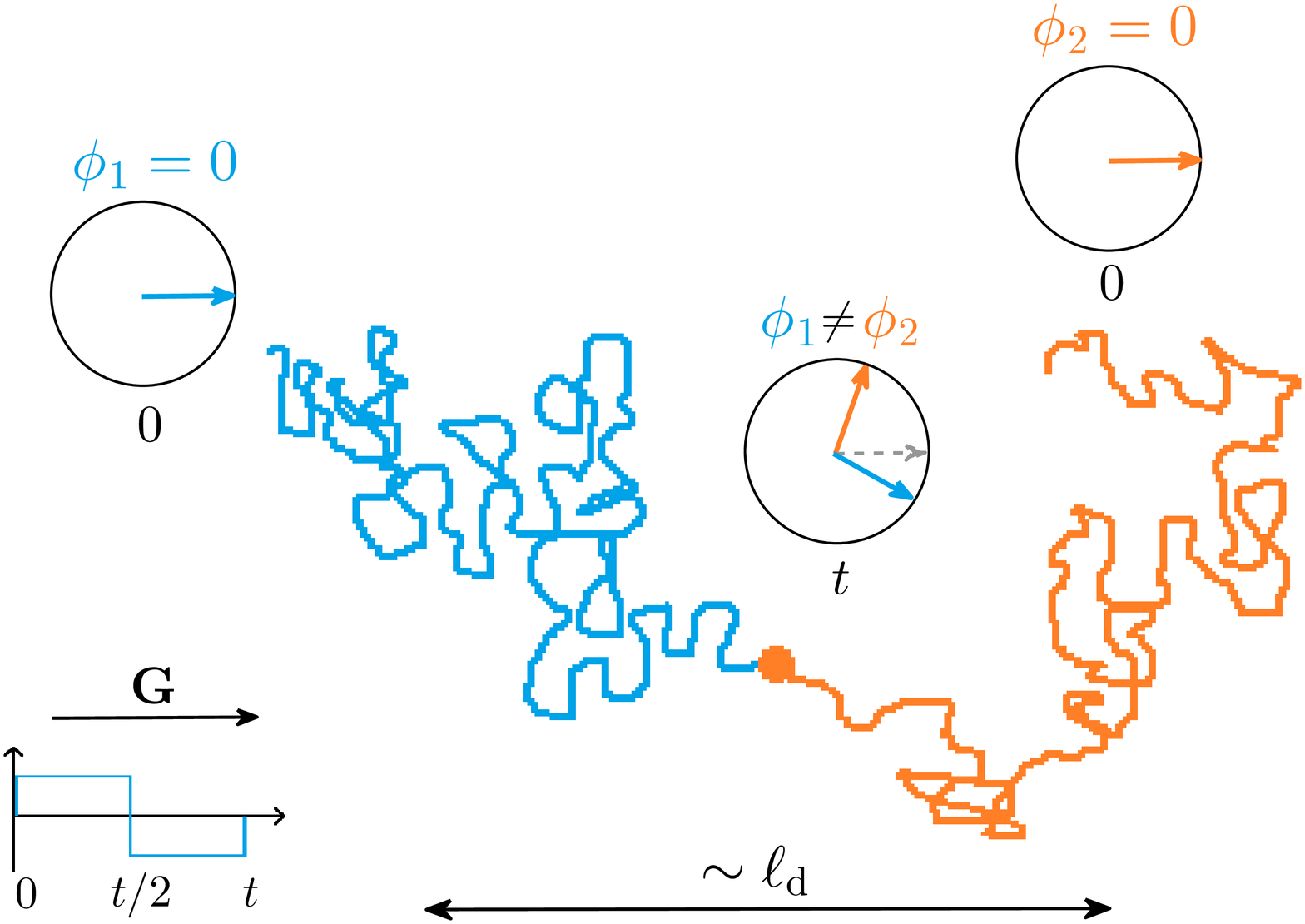} 
\caption{ 
Illustration of diffusion encoding by the gradient. Two particles that
meet each other at position $\mathbf{r}$ at the measurement time $t$
have different histories that lead to different accumulated phases
$\phi_1$ and $\phi_2$ (see Eq. \eqref{eq:phi_brownian}).  In turn, the
resulting phase dispersion leads to signal decay as
$S=\mathbb{E}[e^{i\phi}]$.}
\label{fig:diffusion_encoding}
\end{figure}

Let us consider two particles that meet each other at time $t$ at
position $\mathbf{r}$.  Therefore, they are initially spaced by a
distance of the order of the diffusion length $\ell_{\mathrm d} =
\sqrt{D_0 t}$ (see Fig. \ref{fig:diffusion_encoding}).  We assume that
the pore diameter, $\ell_{\mathrm s}$, is much larger than this
distance, i.e. $\ell_{\mathrm d} \ll \ell_{\mathrm s}$, and that the
particles diffuse far away from the boundaries of the medium so that
we neglect their influence for clarity.  Diffusing in a magnetic field
gradient $\mathbf{g}(t)$ up to time $t$, a particle acquires the
random phase
\begin{equation}
\label{eq:phi_brownian}
\phi = \int_0^t \gamma \, \bigl(\mathbf{g}(t')\cdot \mathbf{r}(t')\bigr)  \,\mathrm{d}t'  ,
\end{equation}
where $\mathbf{r}(t')$ is a random trajectory of the particle.  Let
$\phi_1$ and $\phi_2$ denote two realizations of the random phase
$\phi$ acquired by two particles.
Under a constant gradient amplitude $G = \gamma g$, the random phase
difference accumulated by these two particles until they meet is of
the order of
\begin{equation}
\left|\phi_1 - \phi_2\right| \sim Gt\ell_{\mathrm d} = (\ell_{\mathrm d}/\ell_g)^3\;,
\end{equation}
where $\ell_g = D_0^{1/3} G^{-1/3}$ is the gradient length.
Equivalently, the variance of $\phi$ at position $\mathbf{r}$ scales
as
\begin{equation}
\mathbb{V}[\phi|\mathbf{r}_t = \mathbf{r}] \sim D_0 G^2 t^3 = (\ell_{\mathrm d}/\ell_g)^6\;.
\label{eq:phi_variance_scaling}
\end{equation}
This quantity describes the phase dispersion at a given position, even
if it is position-independent because of the hypothesis of negligible
influence of boundaries.  If $\ell_{\mathrm d} \ll \ell_g$ the typical
phase difference is small so that the spins have strongly correlated
phases.  In other words, two different trajectories yield close values
of $\phi$ and we call this situation ``weak diffusion encoding''.  In
contrast, if $\ell_{\mathrm d} \gtrsim \ell_g$, the typical phase
difference is large and the spins have almost uncorrelated
phases. This is the opposite regime of ``strong diffusion encoding''
where two different trajectories yield very different values of
$\phi$.
Therefore, one can interpret the gradient length $\ell_g$ as the
typical length traveled by particles under the gradient $G$ before
they have decorrelated phases with other spins at the same position,
provided that they do not reach any boundary.

The above reasoning is still valid if the gradient profile is made of
two extended-gradient pulses with no diffusion time in-between them,
such as the profile shown on Fig. \ref{fig:PGSE} with $\Delta =
\delta$.  Indeed, the first (positive) gradient pulse induces a
stronger dephasing than the second (negative) one because particles
are further apart during the first pulse.  However, the argument fails
if the pulses are separated by a diffusion time that is significantly
longer than the duration of the pulses ($\Delta \gg
\delta$).  Indeed, the diffusion step with no gradient mixes particles
from different areas and thus increases dephasing between spins at a
given position.  This is especially the case in the narrow-gradient
regime where the length $\ell_q$, that we describe below, provides
more insight into the formation of the signal.

\subsection*{Phase pattern period $\ell_q$}
\label{section:phase_pattern_period}

The gradient length is an effective way of quantifying the dephasing
acquired by diffusing particles because of their random motion, in
other words, the variance of $\phi$.  In contrast, let us consider the
average phase at a given position $\mathbf{r}$ after the first
gradient pulse of amplitude $G$ and duration $\delta$, such as on
Fig. \ref{fig:PGSE}.  We assume again that the effect of boundaries
can be neglected.  As we consider a single constant-gradient pulse,
the average value of the random phase $\phi$ is not zero and can
evaluated as
\begin{equation}
\mathbb{E}[\phi] = \mathbb{E} \left[\int_0^\delta  \bigl(\mathbf{G}(t) \cdot \mathbf{r}(t)\bigr)\,\mathrm{d}t\right]
= \bigl(\mathbf{G}\cdot \mathbf{r}_0\bigr) \delta\;,
\end{equation}
where $\mathbf{r}_0$ is the starting point.  This implies that the
gradient pulse produces a phase pattern with wavevector $\mathbf{q}$
or equivalently with period $\ell_q$ (up to a $2\pi$ factor):
\begin{equation}
\mathbf{q}=\delta \mathbf{G}\;, \quad \mathrm{and} \quad \ell_q = |\mathbf{q}|^{-1}\;.
\end{equation}
Note that in addition to this phase pattern, one should take into
account the random dephasing computed above that attenuates the
magnetization during the gradient pulse.

Let us consider the limit of infinitely narrow-gradient pulses:
$\delta \to 0$ and $\mathbf{q}$ is constant.  The above estimation
\eqref{eq:phi_variance_scaling} of the variance of $\phi$ just after
the pulse shows that it tends to zero in that limit, therefore the
effect of a narrow-gradient pulse is simply to multiply the
magnetization by the phase pattern $e^{i\mathbf{q}\cdot \mathbf{r}}$
with no attenuation.  In other words, the attenuation of the
magnetization is solely caused by the subsequent diffusion step of
duration $\Delta$ that ``blurs'' the phase pattern of period $\ell_q$.


\begin{thebibliography}{10}

\bibitem{Hahn1950a}		E.~L.~Hahn,
				Spin Echoes,
				Phys. Rev. 80 (1950) 580--594.

\bibitem{Callaghan1991a}	P.~T. Callaghan, 
				{\em Principles of Nuclear Magnetic Resonance Microscopy}
				(Clarendon Press, 1st~ed., 1991).



\bibitem{Price2009a}		W.~Price, 
				{\em NMR studies of translational motion: Principles and applications}
				(Cambridge Molecular Science, 2009).
 
\bibitem{Jones2011a}		D. K. Jones, 
				{\em Diffusion MRI: Theory, Methods, and Applications}
				(Oxford University Press, New York, USA, 2011).


\bibitem{Tuch2003a}		D. S. Tuch, T. G. Reese, M. R. Wiegell and V. J. Wedeen, 
				Diffusion MRI of Complex Neural Architecture, 
				Neuron 40 (2003) 885--895.

\bibitem{Frahm2004a}		J. Frahm, P. Dechent, J. Baudewig and K. D. Merboldt, 
				Advances in functional MRI of the human brain, 
				Prog. Nucl. Magn. Reson. Spectrosc. 44 (2004) 1--32.

\bibitem{LeBihan2012a}		D. Le Bihan and H. Johansen-Berg, 
				Diffusion MRI at 25: Exploring brain tissue structure and function, 
				NeuroImage 61 (2012) 324--341.


\bibitem{Grebenkov2007a}	D.~S. Grebenkov, 
				NMR survey of reflected Brownian motion,
				Rev. Mod. Phys. 79 (2007) 1077--1137.

\bibitem{Kiselev2017a}		V.~G. Kiselev,
				Fundamentals of diffusion MRI physics, 
				NMR Biomed. 30 (2017) e3602.
 
\bibitem{Novikov2018a}		D. S. Novikov, E. Fieremans, S. Jespersen, and V. G. Kiselev,
				Quantifying brain microstructure with diffusion MRI: Theory and parameter estimation,
				NMR Biomed. e3998 (2018).


\bibitem{Torrey1956a}		H.~C.~Torrey, 
				Bloch equations with diffusion terms,
				Phys. Rev. 104 (1956) 563--565.



\bibitem{Stejskal1965a}		E.~O.~Stejskal and J.~E.~Tanner,
				Spin diffusion measurements: Spin echoes in the presence of a time dependent field gradient,
				J. Chem. Phys. 42 (1965) 288--292.

\bibitem{Douglass1958a}		D. C. Douglass and D. W. McCall,
				Diffusion in Paraffin Hydrocarbons,
				J. Phys. Chem. 62 (1958) 1102--1107.


\bibitem{Stoller1991a}		S.~D.~Stoller, W.~Happer, and F.~J.~Dyson, 
				Transverse spin relaxation in inhomogeneous magnetic fields,
				Phys. Rev. A 44 (1991) 7459--7477.


\bibitem{Woessner1963a}		D. E. Woessner,
				NMR spin-echo self-diffusion measurements on fluids undergoing restricted diffusion,
				J. Phys. Chem. 67 (1963) 1365--1367.


\bibitem{Wayne1966a}		R.~C.~Wayne and R.~M.~Cotts, 
				Nuclear-magnetic-resonance study of self-diffusion in a bounded medium,
				Phys. Rev. 151 (1966) 264--272.

\bibitem{vanBeek2004a}		E. J. R. van Beek, J. M. Wild, H.-U. Kauczor, W. Schreiber, J. P. Mugler and E. E. de Lange, 
				Functional MRI of the Lung Using Hyperpolarized 3-Helium Gas, 
				J. Magn. Reson. Imaging 20 (2004) 540--554.



  
  


\bibitem{Mitra1992a}		P.~P.~Mitra, P.~N.~Sen, L.~M.~Schwartz, and P.~Le Doussal,
				Diffusion propagator as a probe of the structure of porous media,
				Phys. Rev. Lett. 68 (1992) 3555. 

\bibitem{Mitra1993a}		P.~P.~Mitra, P.~N.~Sen, and L.~M.~Schwartz,
				Short-time behavior of the diffusion coefficient as a geometrical probe of porous media,
				Phys. Rev. B 47 (1993) 8565.

\bibitem{Moutal2019a}		N. Moutal, I. Maximov, and D. S. Grebenkov, 
				Probing surface-to-volume ratio of an anisotropic medium by diffusion NMR with general gradient encoding, 
				IEEE Trans. Med. Imag. 38 (2019) 2507--2522. 



\bibitem{Robertson1966a}	B.~Robertson, 
				Spin-echo decay of spins diffusing in a bounded region,
				Phys. Rev. 151 (1966) 273--277.

\bibitem{Neuman1974a}		C.~H.~Neuman,
				Spin echo of spins diffusing in a bounded medium,
				J. Chem. Phys. 60 (1974) 4508--4511.

\bibitem{Novikov2011a}		D.~S.~Novikov, E.~Fieremans, J.~H.~Jensen, and J.~A.~Helpern,
				Random walks with barriers,
				Nat. Phys. 7 (2011) 508--514.

\bibitem{Novikov2014a}		D.~S.~Novikov, J.~H.~Jensen, J.~A.~Helpern, and E.~Fieremans,
				Revealing mesoscopic structural universality with diffusion,
				Proc. Nat. Acad. Sci. USA 111 (2014) 5088--5093.


\bibitem{Basser2002a}		P. J. Basser and D. K. Jones,
				Diffusion-tensor MRI: theory, experimental design and data analysis-a technical review,
				NMR Biomed. 15 (2002) 456-467.

\bibitem{Grebenkov2010a}	D. S. Grebenkov, 
				Use, Misuse and Abuse of Apparent Diffusion Coefficients, 
				Concepts Magn. Reson. A 36 (2010) 24--35.





\bibitem{Niendorf1996a}		T. Niendorf, R. M. Dijkhuizen, D. S. Norris, M. van Lookeren Campagne and K. Nicolay, 
				Biexponential diffusion attenuation in various states of brain tissue: implications for 
				diffusion-weighted imaging, 
				Magn. Reson. Med. 36 (1996) 847--857.

\bibitem{Mulkern1999a}		R. V. Mulkern, H. Gudbjartsson, C. F. Westin, H. P. Zengingonul, W. Gartner, C. R. Guttmann, et al., 
				Multicomponent apparent diffusion coefficients in human brain, 
				NMR Biomed. 12 (1999) 51--62.

\bibitem{Clark2000a}		C. A. Clark and D. Le Bihan, 
				Water diffusion compartmentation and anisotropy at high b values in the human brain, 
				Magn. Reson. Med. 44 (2000) 852--859.

\bibitem{Kiselev2007a}		V. G. Kiselev and K. A. Il'yasov, 
				Is the ``biexponential diffusion'' bilexponential?, 
				Magn. Reson. Med. 57 (2007) 464--469.



\bibitem{Karger1985a}		J. K\"arger, 
				NMR self-diffusion studies in heterogeneous systems, 
				Adv. Colloid Interface 23 (1985) 129--148.

\bibitem{Karger1988a}		J. K\"arger, H. Pfeifer and W. Heink, 
				Principles and applications of self-diffusion measurements by nuclear magnetic resonance, 
				in ``Advances in Magnetic Resonance'', ed. J. S. Waugh 
				(Academic Press, San Diego, CA, 1988, vol. 12) 1--89.

\bibitem{Fieremans2010a}	E. Fieremans, D. S. Novikov, J. H. Jensen and J. A. Helpern, 
				Monte Carlo study of a two-compartment exchange model of diffusion, 
				NMR Biomed. 23 (2010) 711--724.

\bibitem{Moutal2018a}		N.~Moutal, M.~Nilsson, D.~Topgaard, and D.~S.~Grebenkov,
				The K\"arger vs bi-exponential model: theoretical insights and experimental validations,
				J. Magn. Reson. 296 (2018) 72--78.


\bibitem{Pfeuffer1999a}		J. Pfeuffer, S. W. Provencher, and R. Gruetter,
				Water diffusion in rat brain in vivo as detected at very large b values is multicompartmental,
				Magn. Reson. Mat. Phys. Biol. Med. 8 (1999) 98--108.


\bibitem{Yablonskiy2003a}	D. A. Yablonskiy, J. L. Bretthorst and J. J. H. Ackerman, 
				Statistical Model for Diffusion Attenuated MR Signal, 
				Magn. Reson. Med. 50 (2003) 664--669.

\bibitem{Callaghan1979a}	P. T. Callaghan, K. W. Jolley, and J. Lelievre,
				Diffusion of water in the endosperm tissue of wheat grains as studied by pulsed field 
				gradient nuclear magnetic resonance,
				Biophys. J. 28 (1979) 133--142.

\bibitem{Yablonskiy2002a}	D. A. Yablonskiy, A. L. Sukstanskii, J. C. Leawoods, D. S. Gierada, G. L. Bretthorst, 
				S. S. Lefrak, J. D. Cooper and M. S. Conradi, 
				Quantitative in vivo Assessment of Lung Microstructure at the Alveolar Level
				with Hyperpolarized 3He Diffusion MRI, 
				Proc. Natl. Acad. Sci. U. S. A. 99 (2002) 3111--3116.


\bibitem{Magin2008a}		R. L. Magin, O. Abdullah, D. Baleanu and X. Joe Zhou, 
				Anomalous diffusion expressed through fractional order differential operators in the Bloch-Torrey equation, 
				J. Magn. Reson. 190 (2008) 255--270.

\bibitem{Palombo2011a}		M. Palombo, A. Gabrielli, S. De Santis, C. Cametti, G. Ruocco and S. Capuani, 
				Spatio-temporal anomalous diffusion in heterogeneous media by nuclear magnetic resonance, 
				J. Chem. Phys. 135 (2011) 034504.


\bibitem{Jensen2005a}		J. H. Jensen, J. A. Helpern, A. Ramani, H. Lu and K. Kaczynski,
				Diffusional kurtosis imaging: the quantification of non-gaussian water diffusion by 
				means of magnetic resonance imaging, 
				Magn. Reson. Med. 53 (2005) 1432--1440.

\bibitem{Trampel2006a}		R. Trampel, J. H. Jensen, R. F. Lee, I. Kamenetskiy, G. McGuinness and G. Johnson, 
				Diffusional Kurtosis imaging in the lung using hyperpolarized 3He, 
				Magn. Reson. Med. 56 (2006), 733--737.

\bibitem{Kiselev2010a}		V. G. Kiselev, 
				The cumulant expansion: an overarching mathematical framework for understanding diffusion NMR, 
				in ``Diffusion MRI: Theory, Methods and Applications,'' ed. D. K. Jones
				(Oxford University Press: Oxford, 2010, ch. 10).

\bibitem{Grebenkov2016a}	D. S. Grebenkov, 
				{\em From the microstructure to diffusion NMR, and back}, 
				in "Diffusion NMR of confined systems", Eds. R. Valiullin (RSC Publishing, Cambridge, 2016).





\bibitem{deSwiet1994a}		T.~M.~de~Swiet and P.~N.~Sen, 
				Decay of nuclear magnetization by bounded diffusion in a constant field gradient, 
				J. Chem. Phys. 100 (1994) 5597--5604.

\bibitem{Hurlimann1995a}	M.~D.~H{\"u}rlimann, K.~G.~Helmer, T.~M.~de~Swiet, and P.~N.~Sen, 
				Spin echoes in a constant gradient and in the presence of simple restriction,
				J. Magn. Reson. A 113 (1995) 260--264.







\bibitem{Grebenkov2014b}	D.~S.~Grebenkov,
				Exploring diffusion across permeable barriers at high  gradients. II. Localization regime,
				J. Magn. Reson. 248 (2014) 164--176.

\bibitem{Grebenkov2017a}	D.~S.~Grebenkov, B.~Helffer, and R.~Henry, 
				The complex Airy operator on the line with a semipermeable barrier,
				SIAM J. Math. Anal. 49 (2017) 1844--1894.

\bibitem{Herberthson2017a}	M.~Herberthson, E.~{\"O}zarslan, H.~Knutsson, and C.-F.~Westin, 
				Dynamics of local magnetization in the eigenbasis of the Bloch-Torrey operator,
				J. Chem. Phys. 146 (2017) 124201.

\bibitem{Grebenkov2018b}	D.~S.~Grebenkov and B.~Helffer, 
				On spectral properties of the Bloch-Torrey operator in two dimensions,
				SIAM J. Math. Anal. 50 (2018) 622--676.

\bibitem{Almog2018a}		Y.~Almog, D.~S.~Grebenkov, and B.~Helffer,
				Spectral semi-classical analysis of a complex Schrödinger operator in exterior domains,
				J. Math. Phys. 59 (2018) 041501.

\bibitem{Almog2019a}		Y.~Almog, D.~S.~Grebenkov, and B.~Helffer,
				On a Schr\"odinger operator with a purely imaginary potential in the semiclassical limit,
				Commun. Part. Diff. Eq. 44 (2019) 1542--1604.

\bibitem{Moutal2019b}		N.~Moutal, K.~Demberg, D.~S.~Grebenkov, and T.~A.~Kuder,
				Localization regime in diffusion NMR: theory and experiments,
				J. Magn. Reson. 305 (2019) 162--174.

\bibitem{Moutal2020a}		N. Moutal, A. Moutal, and D. S. Grebenkov,
				Diffusion NMR in periodic media: efficient computation and spectral properties,
				J. Phys. A: Math. Theor. 53 (2020) 325201.

\bibitem{Grebenkov2018a}	D.~S.~Grebenkov,
				Diffusion MRI/NMR at high gradients: Challenges and perspectives,
				Microporous Mesoporous Mater. 269 (2018) 79--82.





\bibitem{Stejskal1965b}		E. O. Stejskal,
				Use of Spin Echoes in a Pulsed Magnetic-Field Gradient to Study Anisotropic, Restricted Diffusion and Flow,
				J. Chem. Phys. 43 (1965), 3597--3603.

\bibitem{Tanner1968a}		J. E. Tanner and E. O. Stejskal,
				Restricted Self-Diffusion of Protons in Colloidal Systems by the Pulsed-Gradient, Spin-Echo Method,
				J. Chem. Phys. 49 (1968), 1768--1777.


\bibitem{Carr1954a}		H.~Y.~Carr and E.~M.~Purcell,
				Effects of diffusion on free precession in nuclear magnetic resonance experiments,
				Phys. Rev. 94 (1954) 630--638.




\bibitem{Song2000a}		Y.-Q. Song, S. Ryu, and P. N. Sen,
				Determining multiple length scales in rocks,
				Nature 406 (2000) 178--182.


\bibitem{Grebenkov2008a}	D. S. Grebenkov, 
				Laplacian Eigenfunctions in NMR I. A Numerical Tool, 
				Conc. Magn. Reson. 32A (2008), 277--301. 



\bibitem{Swiet1995a}		T.~M.~de.~Swiet,
				Diffusive Edge Enhancement in Imaging,
				J. Magn. Reson. B 109 (1995) 12--18.


\bibitem{LeDoussal1992a}	P.~Le~Doussal and P.~N.~Sen,
				Decay of nuclear magnetization by diffusion in a parabolic magnetic field: An exactly solvable model,
				Phys. Rev. B 46 (1992) 3465--3485.



\bibitem{Novikov2010a}		D. S. Novikov and V. G. Kiselev, 
				Effective medium theory of a diffusion weighted signal, 
				NMR Biomed. 23 (2010) 682--697.



\bibitem{Axelrod2001a}		S. Axelrod and P. N. Sen,
				Nuclear magnetic resonance spin echoes for restricted diffusion in an inhomogeneous field: Methods and asymptotic regimes,
				J. Chem. Phys. 114 (2001) 6878--6895.



\bibitem{Brownstein1979a}	K. R. Brownstein and C. E. Tarr,
				Importance of Classical Diffusion in NMR Studies of Water in Biological Cells,
				Phys. Rev. A 19 (1979) 2446--2453.



\bibitem{Kuder2013a}		T. A. Kuder, P. Bachert, J. Windschuh and F. B. Laun, 
				Diffusion Pore Imaging by Hyperpolarized Xenon-129 Nuclear Magnetic Resonance,
				Phys. Rev. Lett. 111 (2013) 028101.

\bibitem{Hertel2013a}		S. Hertel, M. Hunter and P. Galvosas, 
				Magnetic resonance pore imaging, a tool for porous media research, 
				Phys. Rev. E 87 (2013) 030802.



\bibitem{Eriksson2013a}		S. Eriksson, S. Lasic, and D. Topgaard,
				Isotropic diffusion weighting in PGSE NMR by magic-angle spinning of the q-vector,
				J. Magn. Reson 226 (2013), 13--18.

\bibitem{Topgaard2017a}		D. Topgaard, 
				Multidimensional diffusion MRI, 
				J. Magn. Reson. 275 (2017), 98--113.







\bibitem{Latour1995a}		L. L. Latour, R. L. Kleinberg, P. P. Mitra, and C. H. Sotak,
				Pore-Size Distributions and Tortuosity in Heterogeneous Porous Media,
				J. Magn. Reson. A 112 (1995), 83--91.


\bibitem{Debye1957a}		P.~Debye, H.~R.~Anderson, and H.~Brumberger,
				Scattering by an Inhomogeneous Solid. II. The correlation function and its application,
				J. Appl. Phys. 28 (1957) 679--683.

\bibitem{Sen1995a}		P.~N.~Sen, M.~D.~H{\"u}rlimann, and T.~M.~de~Swiet,
				Debye-Porod law of diffraction for diffusion in porous media,
				Phys. Rev. B 51 (1995) 601--604.

\bibitem{Froehlich2006a}	A.~F.~Fr{\o}hlich, L.~Ostergaard, and V.~G.~Kiselev,
				Effect of impermeable boundaries on diffusion-attenuated MR signal,
				J. Magn. Reson. 179 (2006) 223--233.



\bibitem{Callaghan1991b}	P.~T.~Callaghan, A.~Coy, D.~MacGowan, K.~J.~Packer, and F.~O.~Zelaya,
				Diffraction-like Effects in NMR Diffusion Studies of Fluids in Porous Solids,
				Nature 351 (1991) 467--469.

\bibitem{Callaghan1992a}	P.~T.~Callaghan, A.~Coy, T.~P.~J.~Halpin, D.~MacGowan, K.~J.~Packer, abd F.~O.~Zelaya, 
				Diffusion in porous systems and the influence of pore morphology in pulsed gradient 
				spin-echo nuclear magnetic resonance studies,
				J. Chem. Phys. 97 (1992) 651--662.

\bibitem{Coy1994a}		A.~Coy and P.~T.~Callaghan,
				Pulsed gradient spin echo nuclear magnetic resonance for molecules diffusing between 
				partially reflecting rectangular barriers,
				J. Chem. Phys. 101 (1994) 4599--4609.

\bibitem{Callaghan1995a}	P.~T.~Callaghan, 
				Pulsed-gradient spin-echo NMR for planar, cylindrical, and spherical pores under conditions of wall relaxation,
				J. Magn. Reson. A 113 (1995) 53--59.

\bibitem{Linse1995a}		P.~Linse and O.~S{\"o}derman, 
				The validity of the short-gradient-pulse approximation in NMR studies of restricted diffusion. 
				Simulations of molecules diffusing between planes, in cylinders, and spheres,
				J. Magn. Reson. A 116 (1995) 77--86.


\end{thebibliography}
\end{document}